\documentclass[10pt,a4paper,twoside,dvipdfm]{article}
\usepackage{epsfig}
\usepackage{baltlat6}
\usepackage{array}
\usepackage{wrapfig}
\usepackage{dcolumn}
\begin{document}
\ \
\vspace{0.5mm}
\setcounter{page}{21}
\vspace{8mm}

\titlehead{Baltic Astronomy, vol.\,17, 21--37, 2008}

\titleb{YOUNG STARS IN THE CAMELOPARDALIS DUST AND\\ MOLECULAR CLOUDS.
IV. SPECTRAL OBSERVATIONS OF\\ THE SUSPECTED YSOs}

\begin{authorl}
\authorb{C. J. Corbally}{1} and
\authorb{V. Strai\v zys}{2}
\end{authorl}

\begin{addressl}
\addressb{1}{Vatican Observatory Research Group, Steward Observatory,
Tucson, Arizona 85721, U.S.A.; corbally@as.arizona.edu}

\addressb{2}{Institute of Theoretical Physics and Astronomy, Vilnius
University,\\  Go\v stauto 12, Vilnius LT-01108, Lithuania;
straizys@itpa.lt}
\end{addressl}

\submitb{Received 2008 March 15; accepted 2008 March 30}

\begin{summary} In the first three papers of this series, about 200
objects in Camelopardalis and the nearby areas of Cassiopeia, Perseus
and Auriga were suspected of being pre-main-sequence stars in different
stages of evolution.  To confirm the evolutionary status of the 15
brightest objects, their far-red range (600--950 nm) spectra were
obtained.  Almost all these objects are young stars with emissions in
H$\alpha$, O\,I, Ca\,II and P9 lines.  The equivalent widths of emission
lines and approximate spectral classes of the objects are determined.
\end{summary}


\resthead{Young stars in the Camelopardalis dust and molecular clouds.
IV.}{C. J. Corbally, V. Strai\v zys}

\sectionb{1}{INTRODUCTION}

In the previous papers of this series (Strai\v zys \& Laugalys 2007a,b,
2008, Papers I, II and III) about 250 young stellar objects were
identified in the Camelopardalis segment of the Milky Way, including the
nearby areas of Cassiopeia, Perseus and Auriga ($\ell$, $b$\,=
132--158\degr, $\pm$\,12\degr).  The limits of the area were defined by
the distribution of the Cam OB1 association members.  Paper I listed 43
massive members of the association and 18 stars of lower masses which
were known to exhibit emission in H$\alpha$ or to belong to irregular
variable stars of types IN and IS.  In Paper II, 142 suspected young
stellar objects (YSOs) were selected on the ground of infrared
photometry taken from the 2MASS, IRAS and MSX databases.  In Paper III,
50 suspected YSOs were selected in the same way in a
3\degr\,$\times$\,3\degr\ area covering the densest part of the dust
cloud T\,942 (Dobashi et al. 2005), with the high-mass object GL\,490
embedded.  Although some of the suspected YSOs (especially those without
IRAS and/or MSX measurements) can be stars with dense envelopes
belonging to the asymptotic branch, or even heavily reddened spiral
galaxies, there is a strong evidence of star-forming activity in the
Camelopardalis dust and molecular clouds.

Naturally, the evolutionary status of each suspected YSO needs spectral
confirmation.  For the beginning, spectral observations were obtained
for 15 objects from the described lists, including seven stars for which
the presence of the emission lines was known earlier, and some brightest
stars from the Paper II and Paper III lists.  The list of the observed
stars is presented in Table 1, and their identification charts of the
1.8\arcmin\,$\times$\,1.8\arcmin\ size are given in Figure 1.

\begin{table}[!t]
\begin{center}
\vbox{\small\tabcolsep=6pt
\parbox[c]{120mm}{\baselineskip=10pt
{\normbf\ \ Table 1.}{\norm\ List of the investigated stars.
$F$ is the red photographic magnitude. In the last column, YSO
means the known or suspected young stellar object, H$\alpha$
means the star with H$\alpha$ emission found in objective-prism spectra.
\lstrut}}
\begin{tabular}{lcccrrl}
\tablerule
 Name  &  RA\,(2000) & DEC\,(2000) & $\ell$ & $b$~~~   &  $F$~~   &Type \\
\tablerule
SL\,21    & 2 23 18.4 &  +61 25 41 &   133.688&   +0.490 &  15.14 &  YSO       \\[2pt]
SL\,37    & 2 27 16.0 &  +62 00 50 &   133.919&   +1.204 &  13.31 &  YSO\\[2pt]
SL\,78    & 2 51 47.0 &  +55 42 01 &   139.369&  --3.283 &  13.10 &  YSO\\[2pt]
KW\,14-24 & 3 01 21.6 &  +60 28 56 &   138.280&   +1.541 &  13.40 &  H$\alpha$\\[2pt]
SL\,75    & 3 03 25.8 &  +60 23 10 &   138.550&   +1.580 &  14.18 &  YSO\\[2pt]
SL\,79    & 3 10 46.3 &  +59 30 04 &   139.788&   +1.269 &  15.12 &  YSO\\[2pt]
SL\,82    & 3 17 26.1 &  +60 09 42 &   140.160&   +2.268 &   9.31 &  H$\alpha$\\[2pt]
GL\,490   & 3 27 38.8 &  +58 47 00 &   142.000&   +1.820 &  17.67 &  YSO\\[2pt]
IRAS 03243+5819 & 3 28 14.6 & +58 29 38 & 142.227 & +1.624 & 15.75 & YSO\\[2pt]
SL\,101   & 3 29 07.6 &  +57 01 34 &   143.152&   +0.479 &  14.90 &  YSO\\[2pt]
SL\,158   & 3 30 05.5 &  +58 13 25 &   142.580&   +1.539 &  14.63 &  YSO\\[2pt]
Gahm 25   & 3 54 20.3 &  +53 05 34 &   148.399&  --0.487 &  14.87 &  H$\alpha$\\[2pt]
Gahm 23   & 3 57 55.0 &  +52 25 37 &   149.241&  --0.650 &  14.26 &  H$\alpha$\\[2pt]
Gahm 22   & 3 58 31.7 &  +52 02 42 &   149.560&  --0.880 &  14.21 &  H$\alpha$\\[2pt]
Gahm 21   & 3 59 53.4 &  +51 31 46 &   150.055&  --1.135 &  13.76 &  H$\alpha$ \\
\tablerule
\end{tabular}
}
\end{center}
\end{table}

\sectionb{2}{SPECTRAL OBSERVATIONS}

The spectra were taken with the Bollen \& Chivens spectrograph on the
Steward Observatory 2.3 m telescope at Kitt Peak, using the 400 g/mm
red-blazed grating, giving a resolution of 5.7 \AA\ and a range from
6075 to 9395 \AA\ on the BCSpec 1200\,$\times$\,800 CCD detector. The
slit width was 1.5\arcsec.

The spectra, shown in Figure 2 in a widened form, were reduced using
IRAF software.  The energy distributions, shown in Figure 3 (a--p), were
obtained by calibrating the spectra with the spectrophotometric standard
Hiltner 600 and removing an atmospheric extinction curve typical for
Kitt Peak.  For the classification of stars the criteria described by
Danks \& Dennefeld (1994) in the far-red and near-infrared spectral
region were applied.  The results of classification are only approximate
since the spectral features at this resolution and S/N do not give the
discrimination of luminosities and, in some cases, even of spectral
subclasses.  Some interference is also due to unexcluded H$_2$O and
O$_2$ telluric bands and some spectral fringing which was not
eliminated completely for the low-S/N spectra.
  The emission line intensities of the spectral classifications are
scaled in relation to those given by Herbig (1962, Table I) for the T
Tauri stars CW Tau, CY Tau, BP Tau and T Tau, whose spectra were
observed in the same nights as the program stars.


\vbox{
\centerline{\hbox{
\psfig{figure=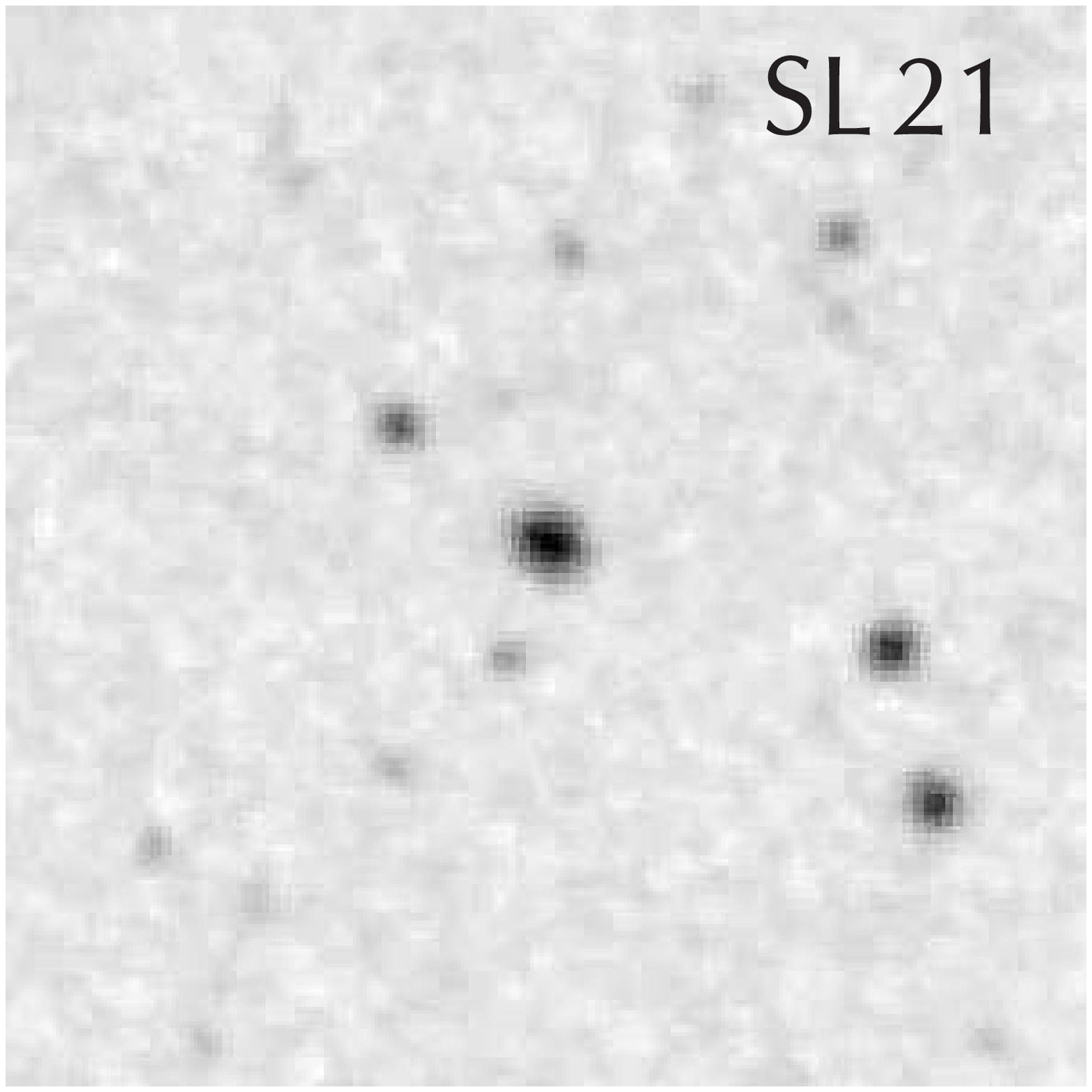,width=35mm,angle=0,clip=true}
\hskip2mm
\psfig{figure=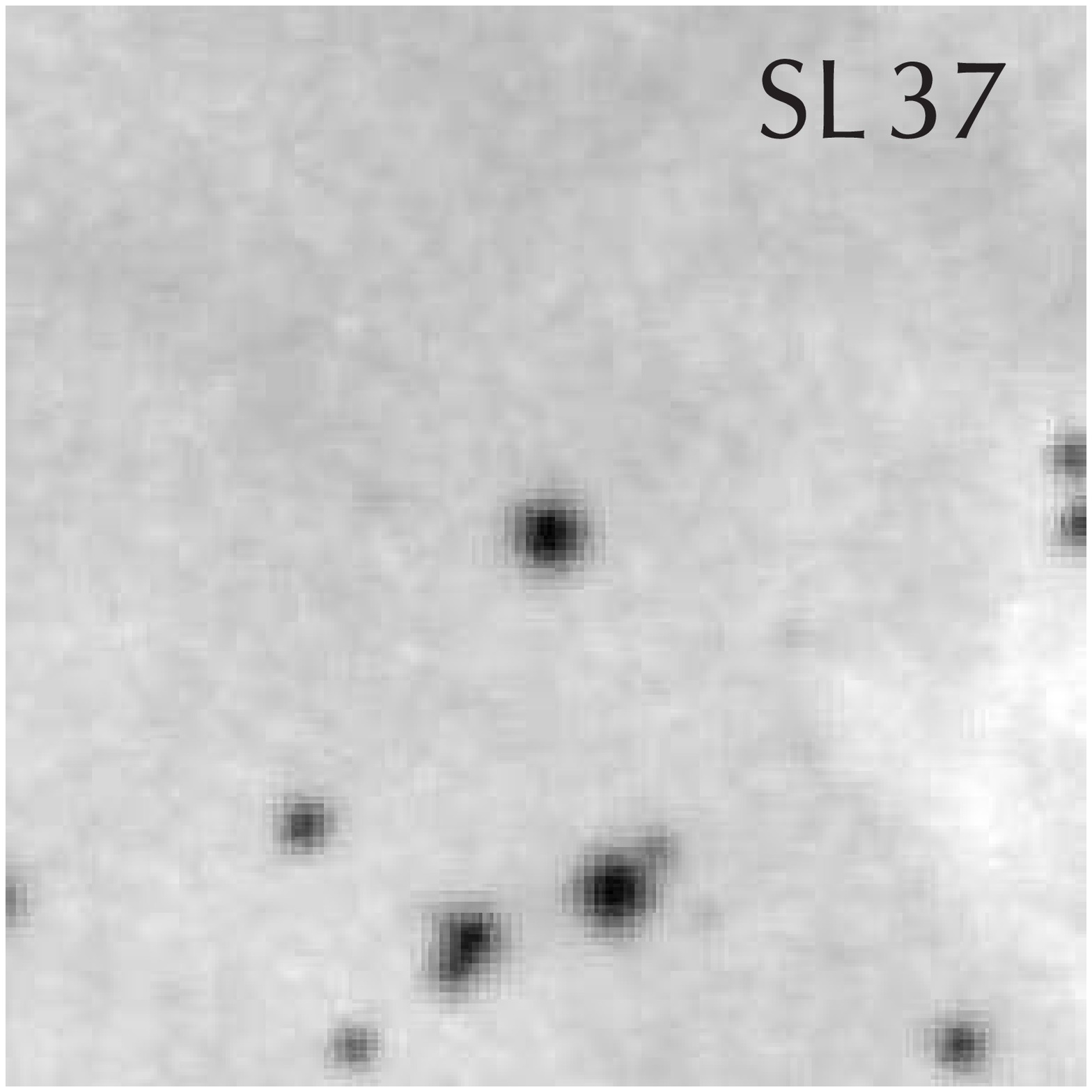,width=35mm,angle=0,clip=true}
\hskip2mm
\psfig{figure=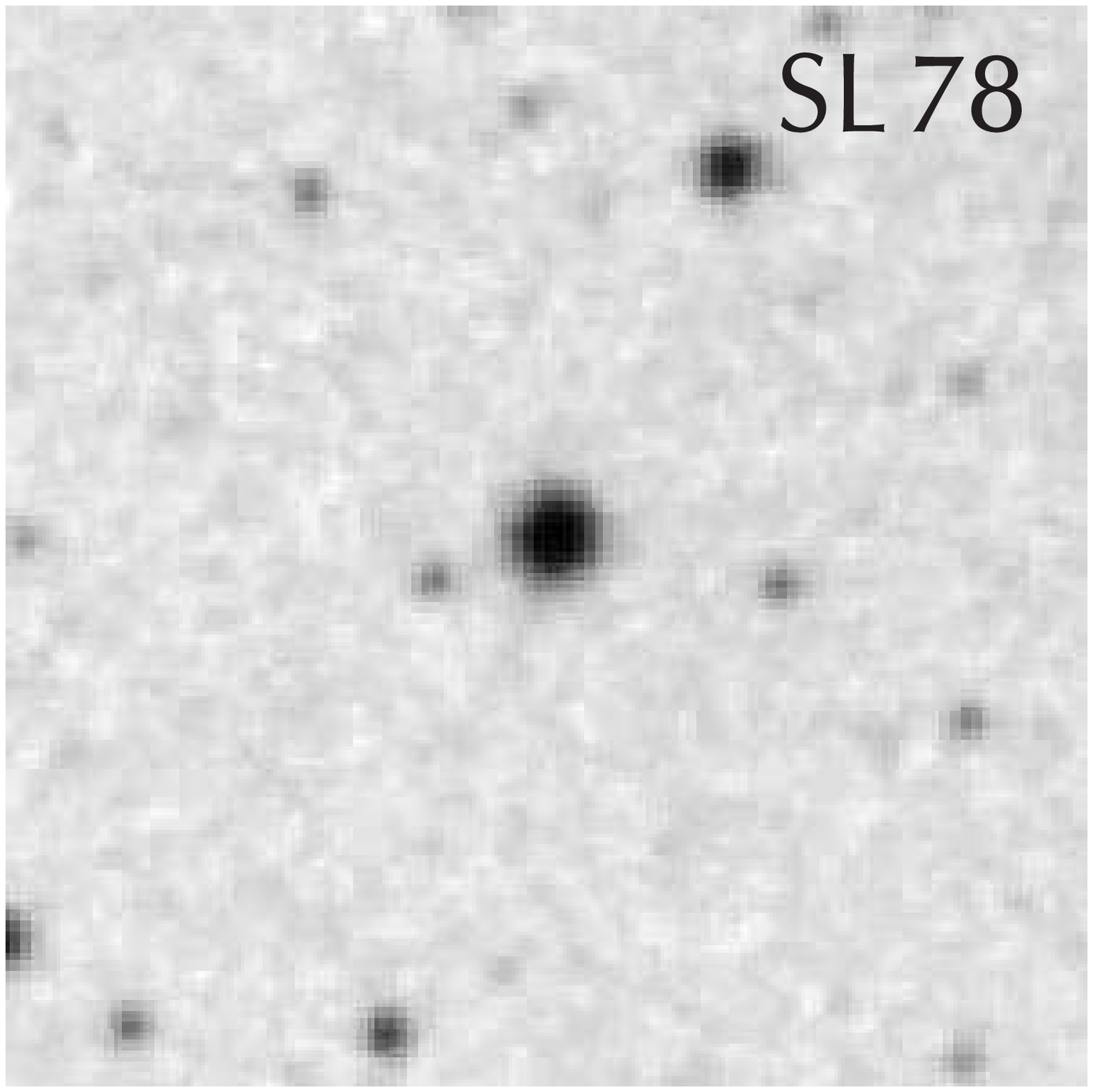,width=35mm,angle=0,clip=true}
}}
\vskip2mm
\centerline{\hbox{
\psfig{figure=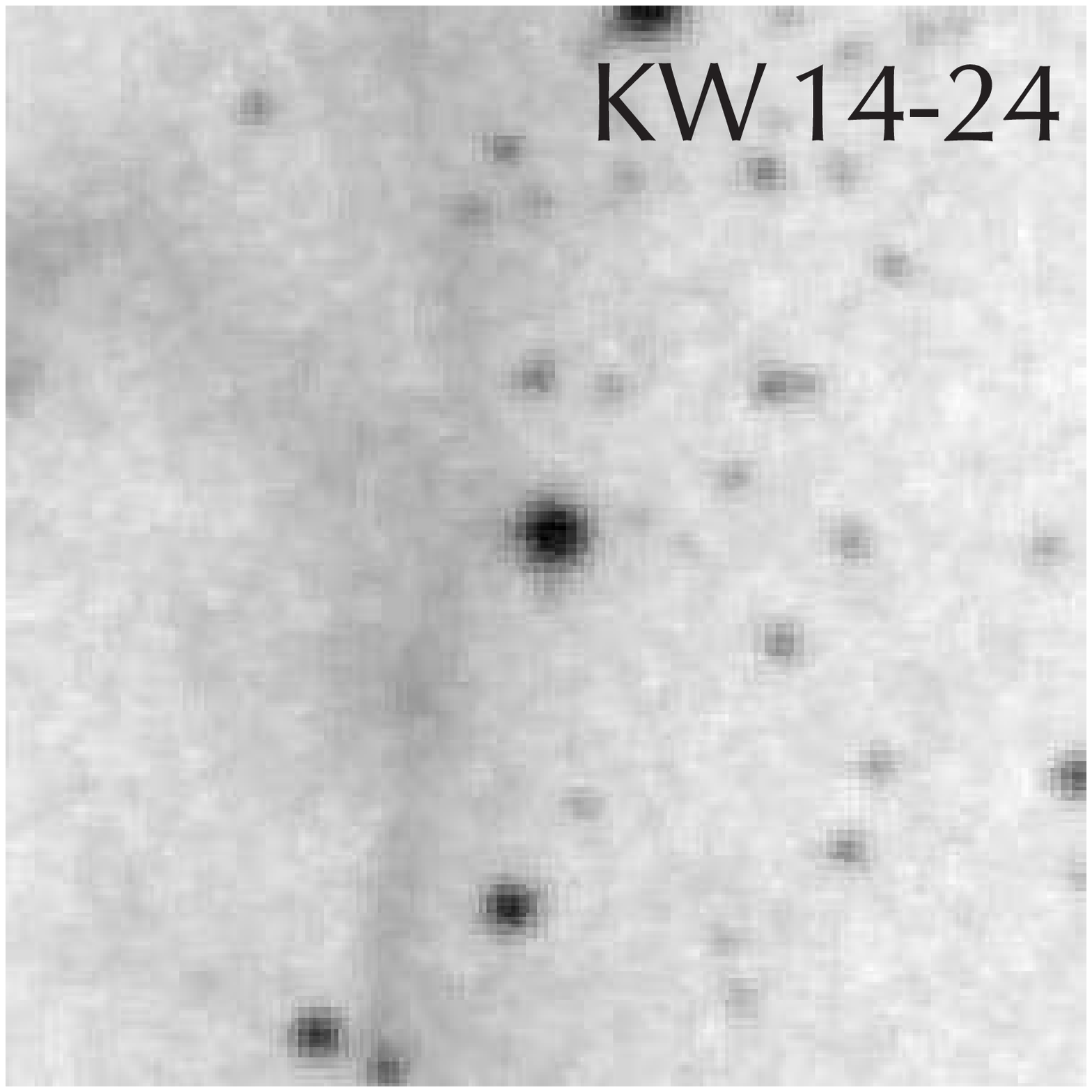,width=35mm,angle=0,clip=true}
\hskip2mm
\psfig{figure=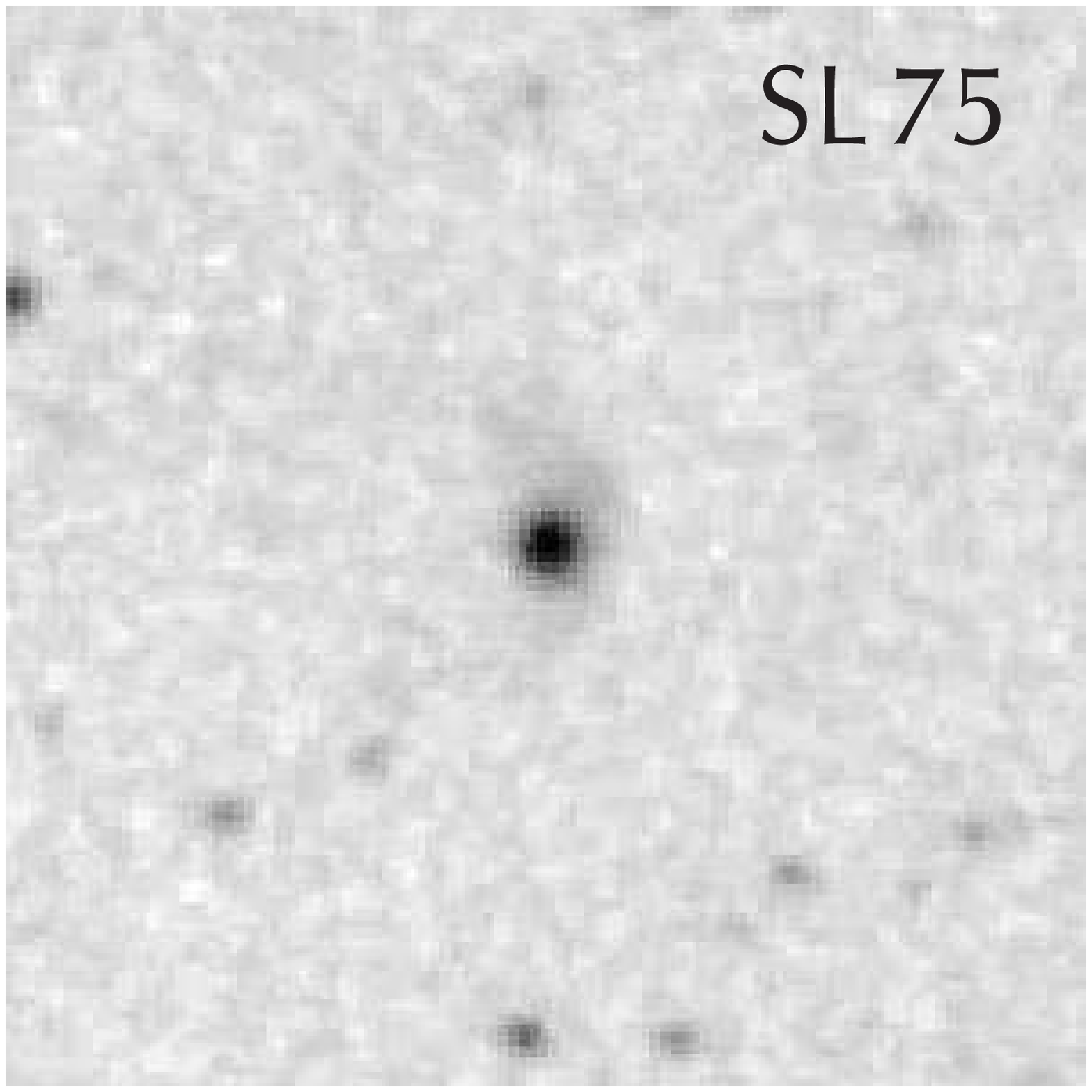,width=35mm,angle=0,clip=true}
\hskip2mm
\psfig{figure=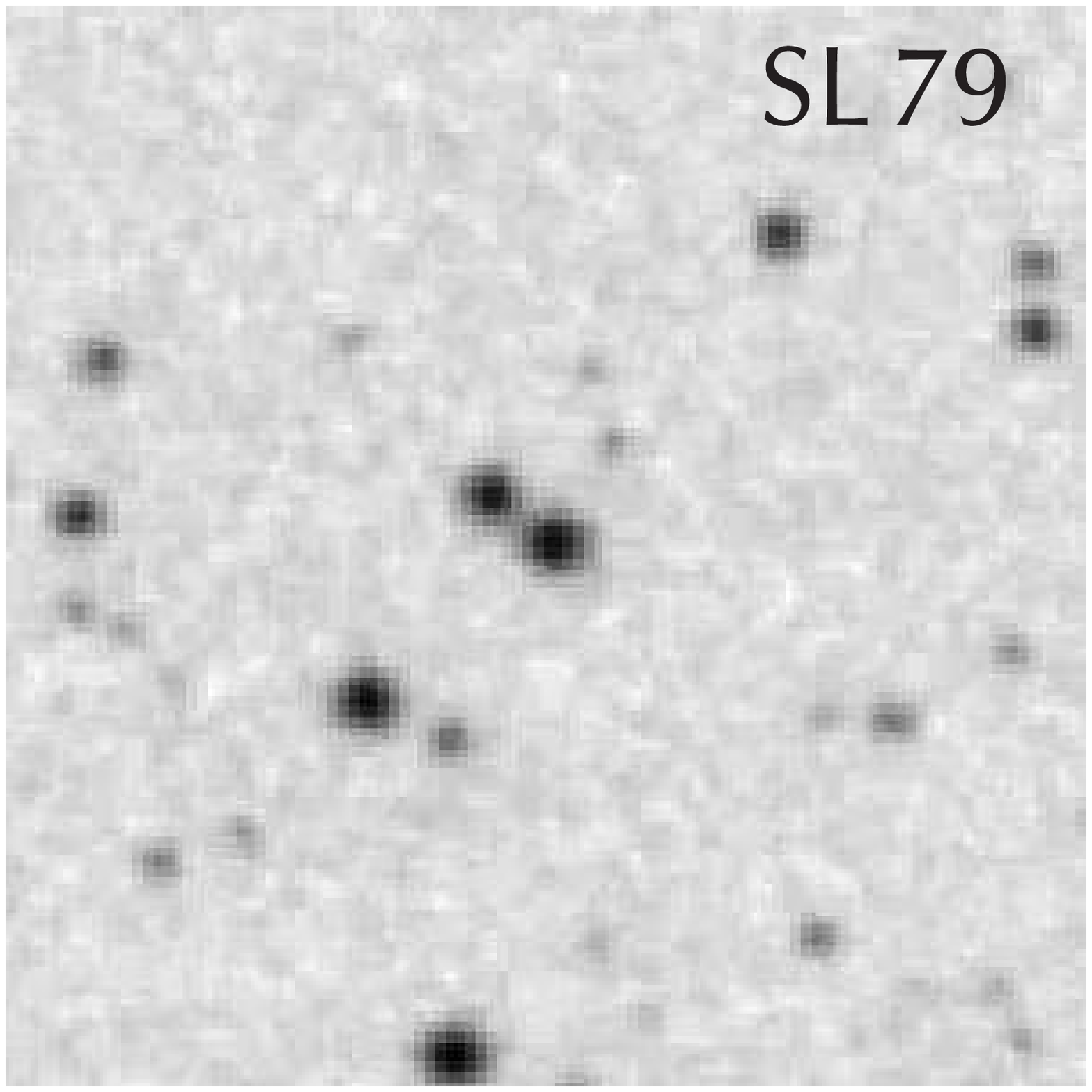,width=35mm,angle=0,clip=true}
}}
\vskip2mm
\centerline{\hbox{
\psfig{figure=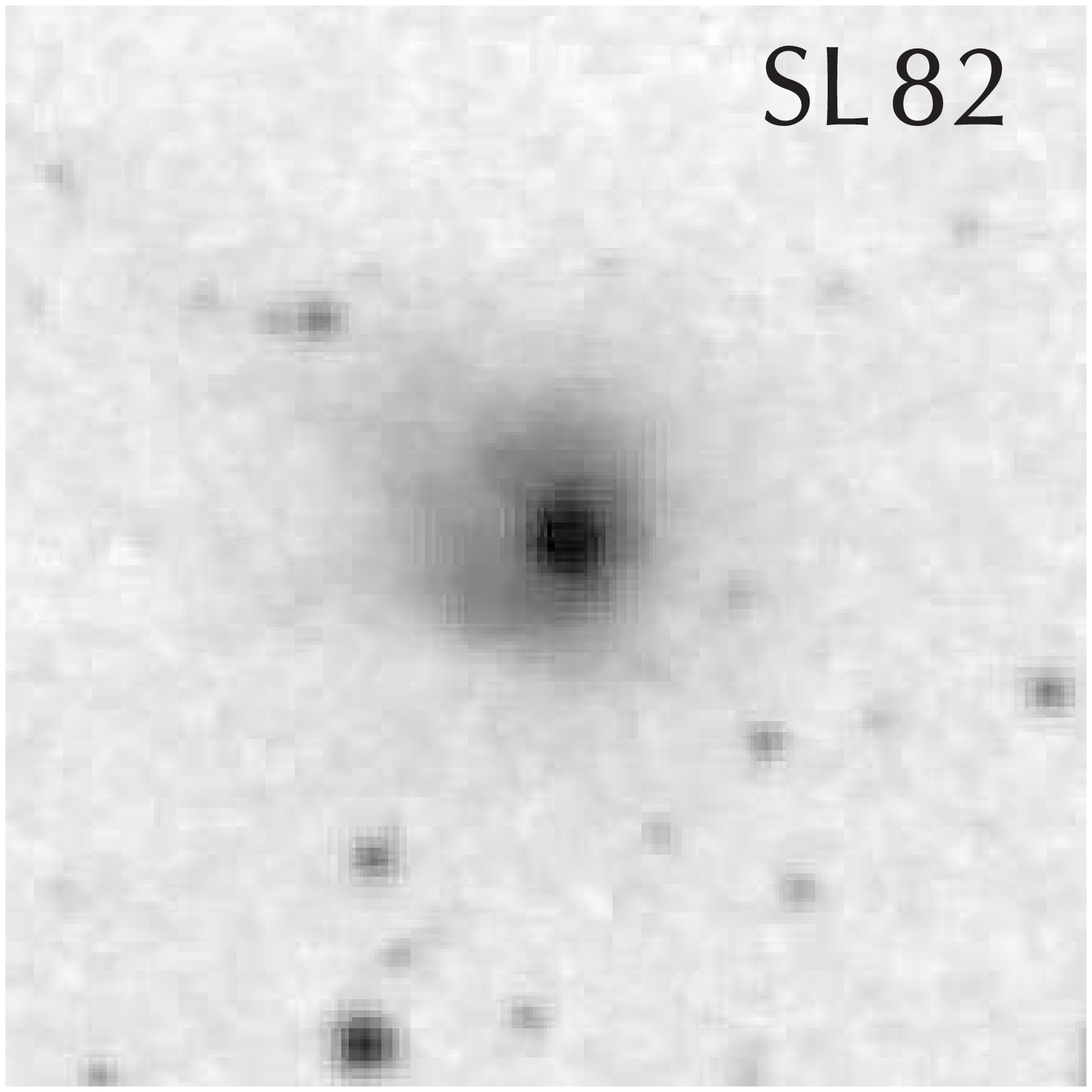,width=35mm,angle=0,clip=true}
\hskip2mm
\psfig{figure=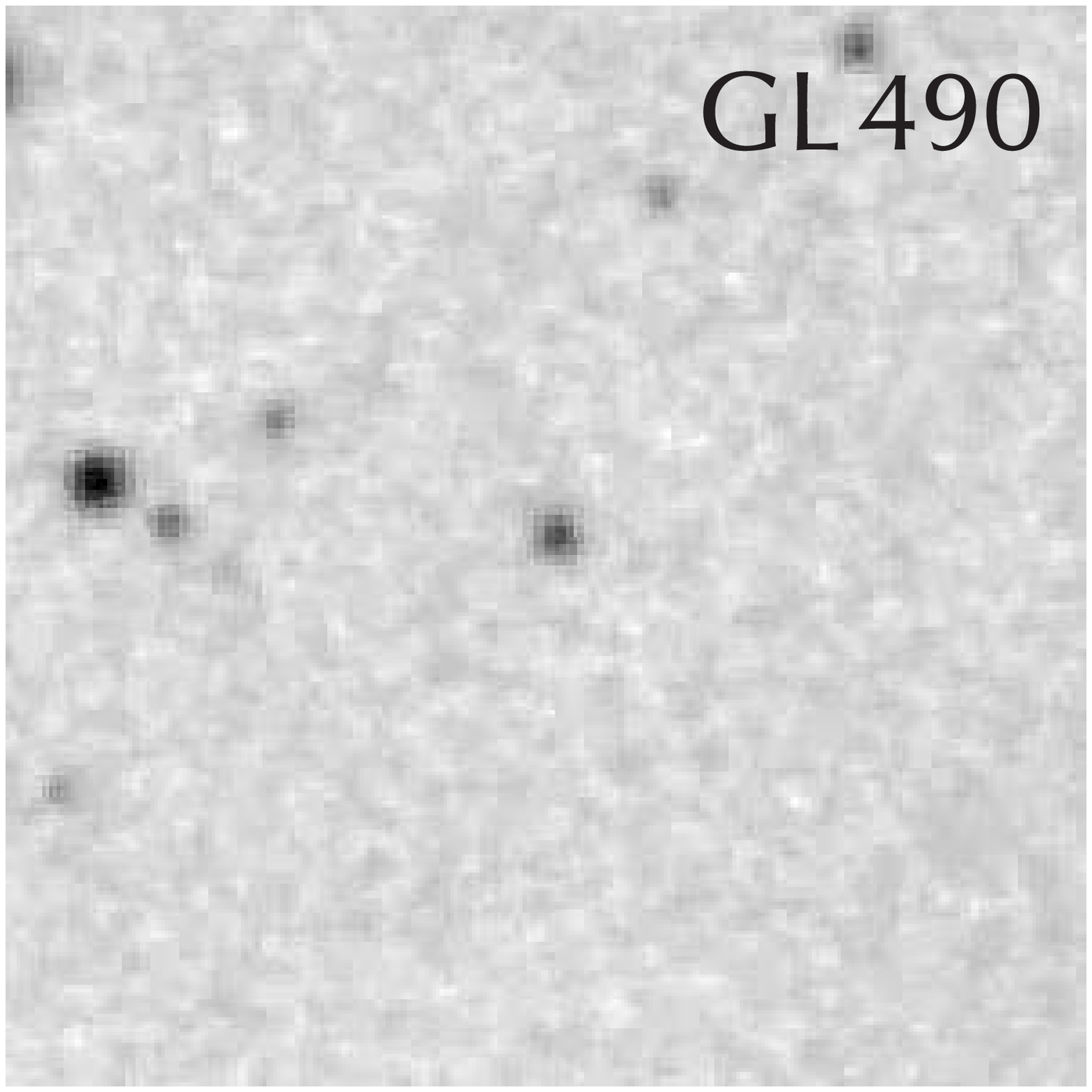,width=35mm,angle=0,clip=true}
\hskip2mm
\psfig{figure=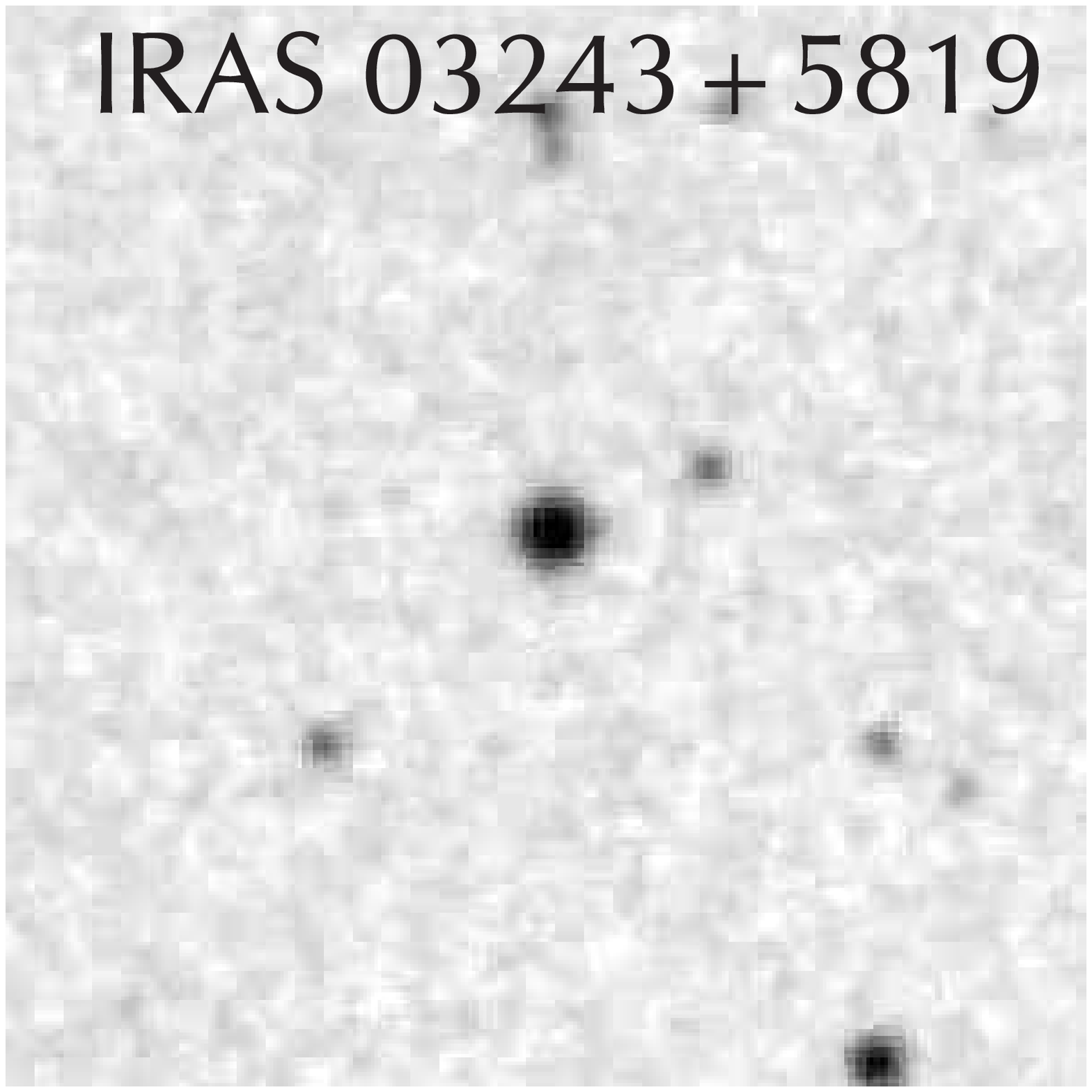,width=35mm,angle=0,clip=true}
}}
\vskip2mm
\centerline{\hbox{
\psfig{figure=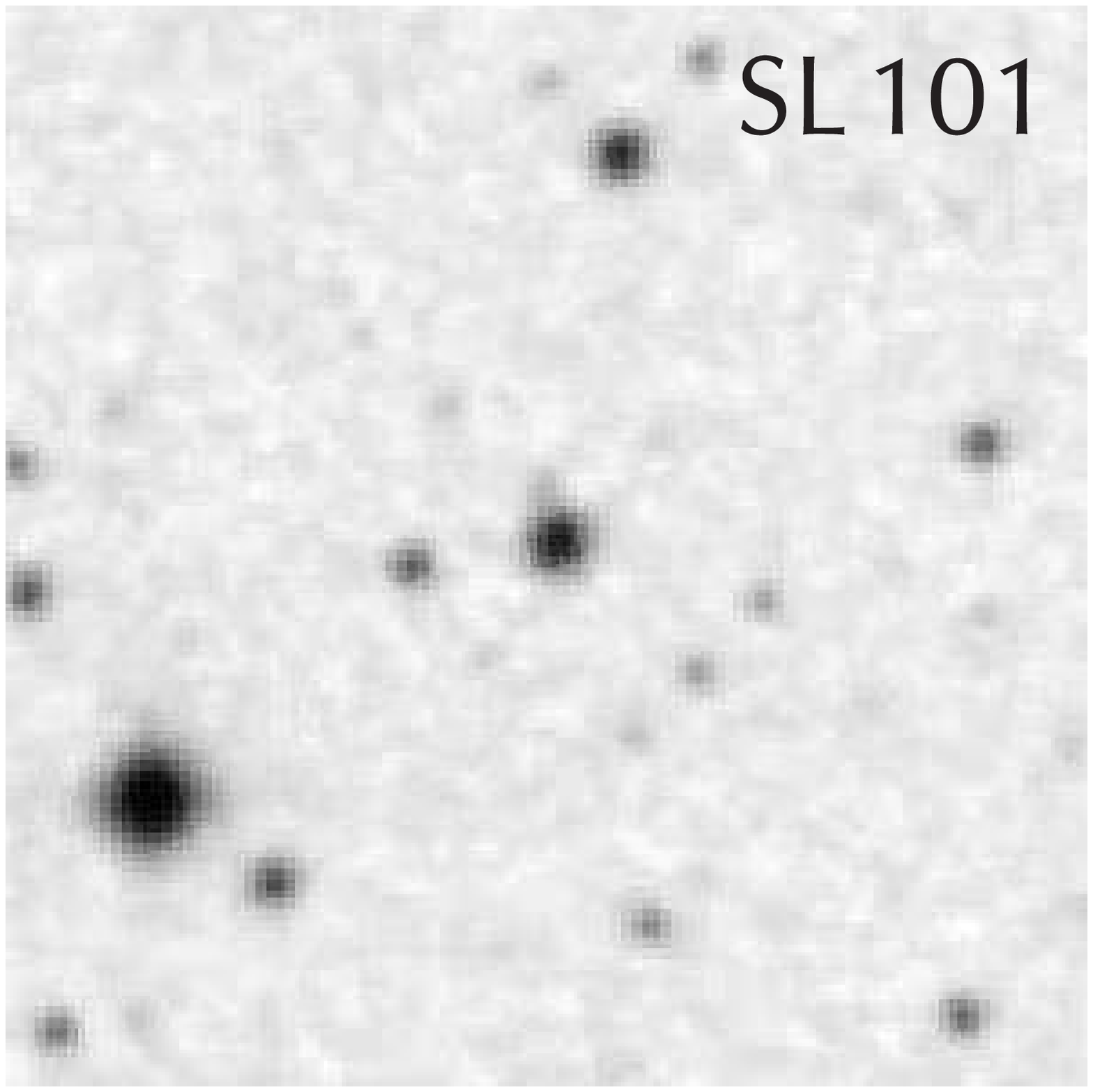,width=35mm,angle=0,clip=true}
\hskip2mm
\psfig{figure=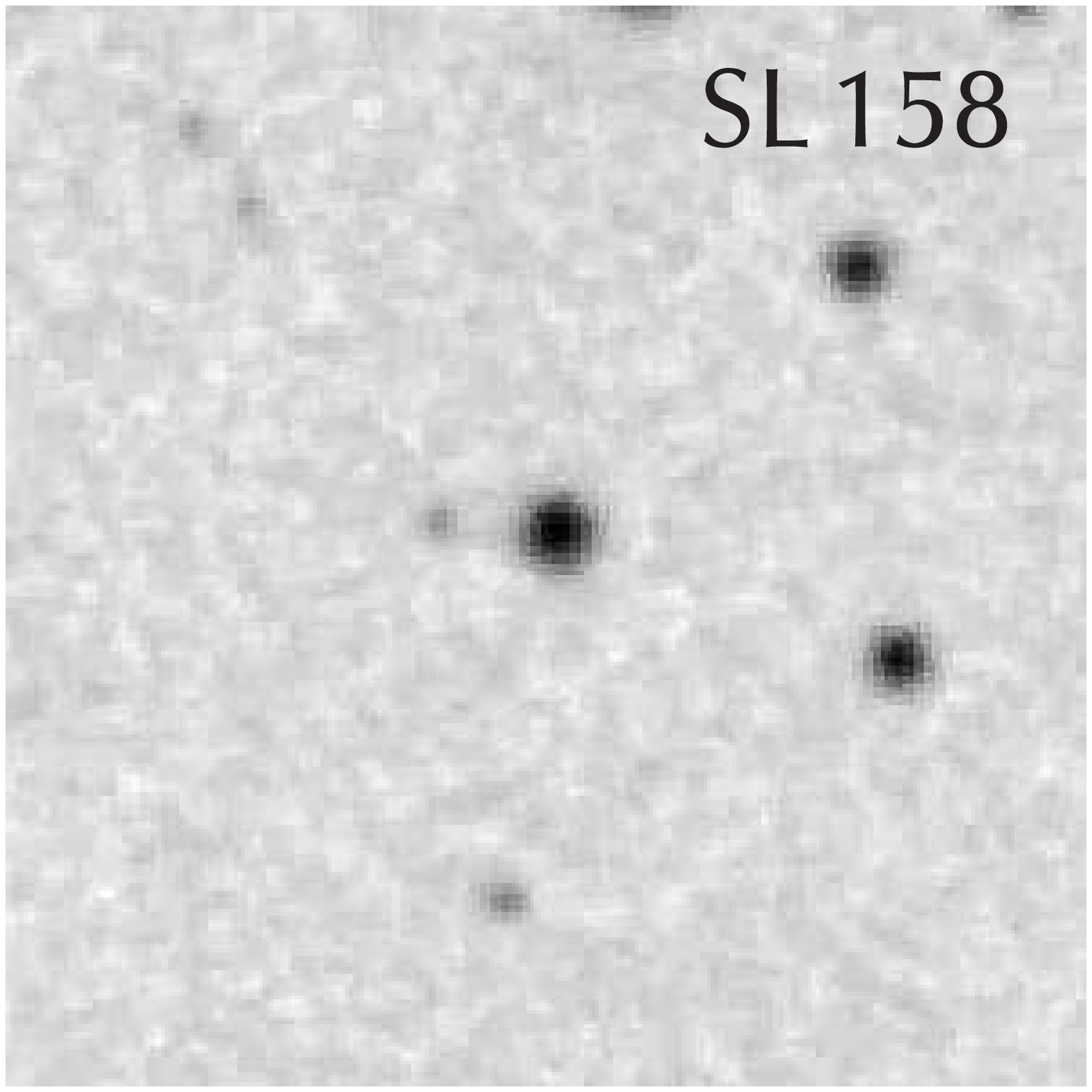,width=35mm,angle=0,clip=true}
\hskip2mm
\psfig{figure=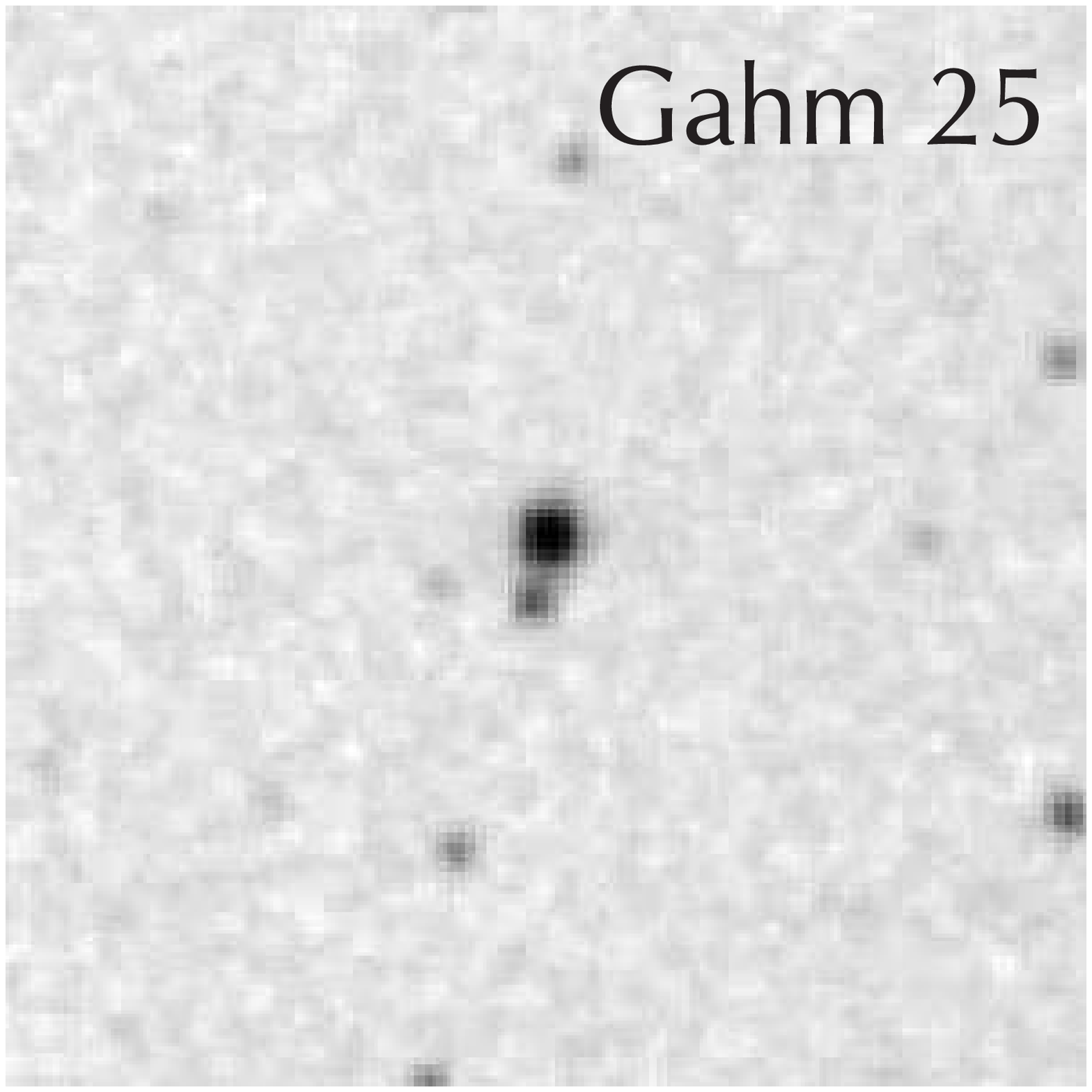,width=35mm,angle=0,clip=true}
}}
\vskip2mm
\centerline{\hbox{
\psfig{figure=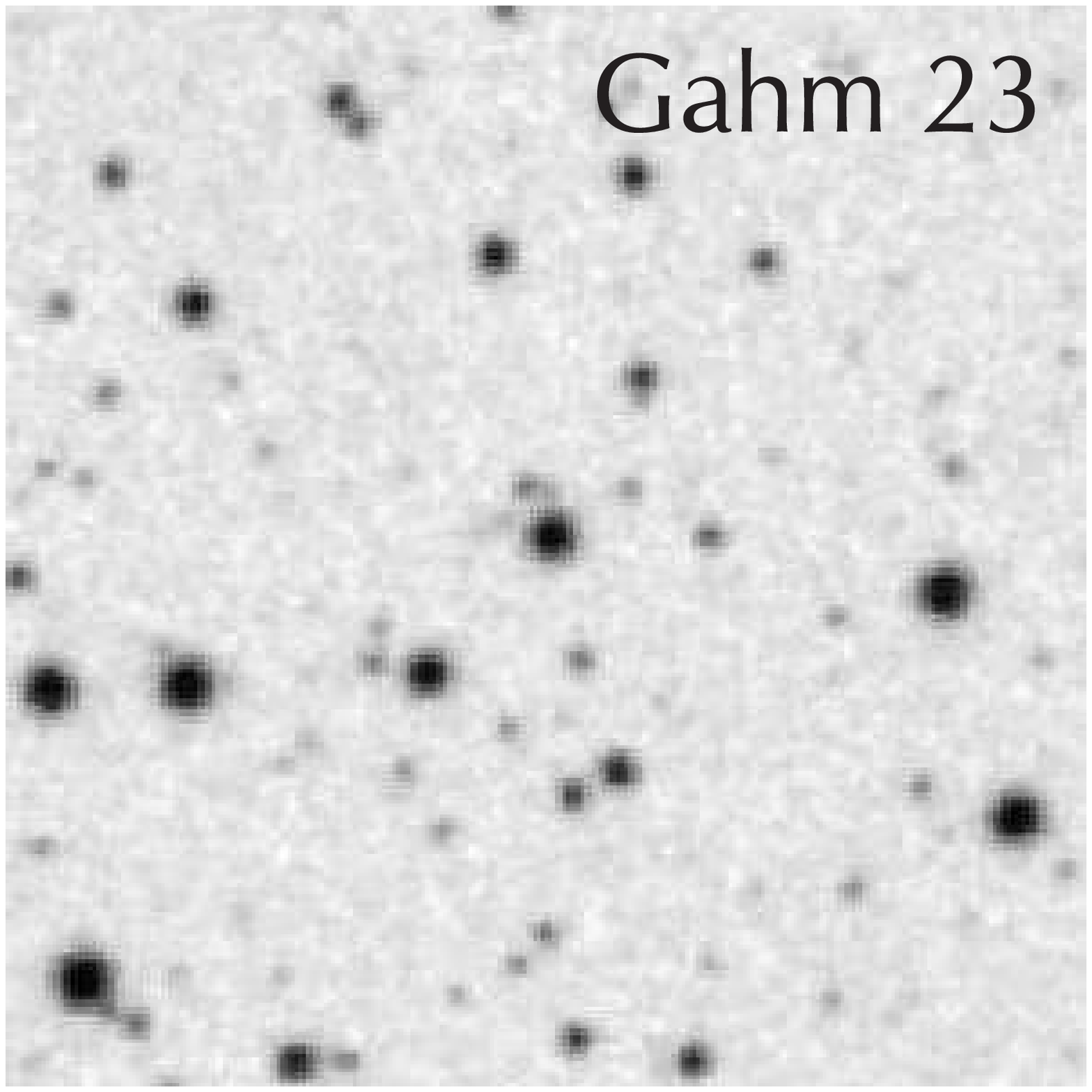,width=35mm,angle=0,clip=true}
\hskip2mm
\psfig{figure=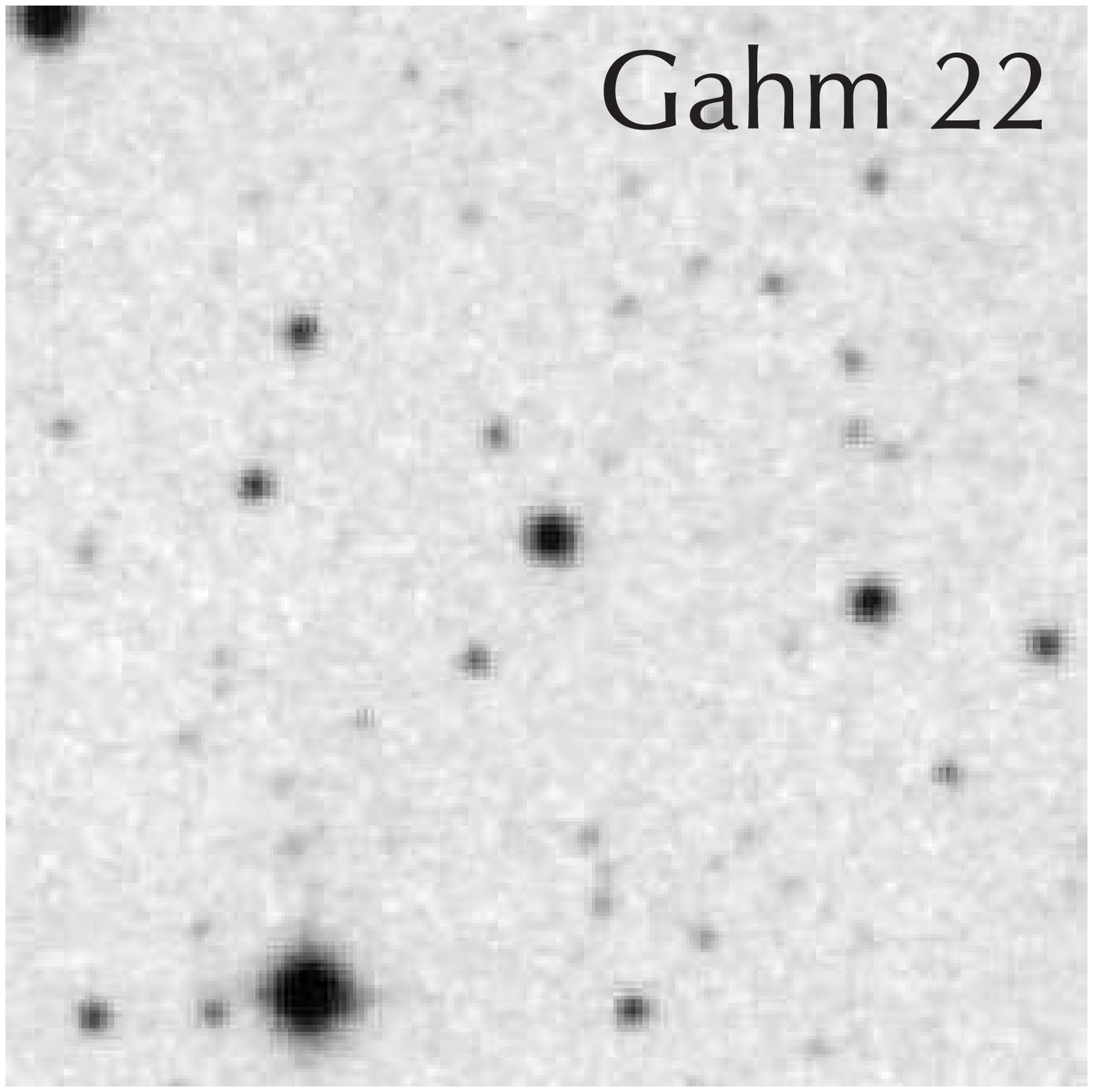,width=35mm,angle=0,clip=true}
\hskip2mm
\psfig{figure=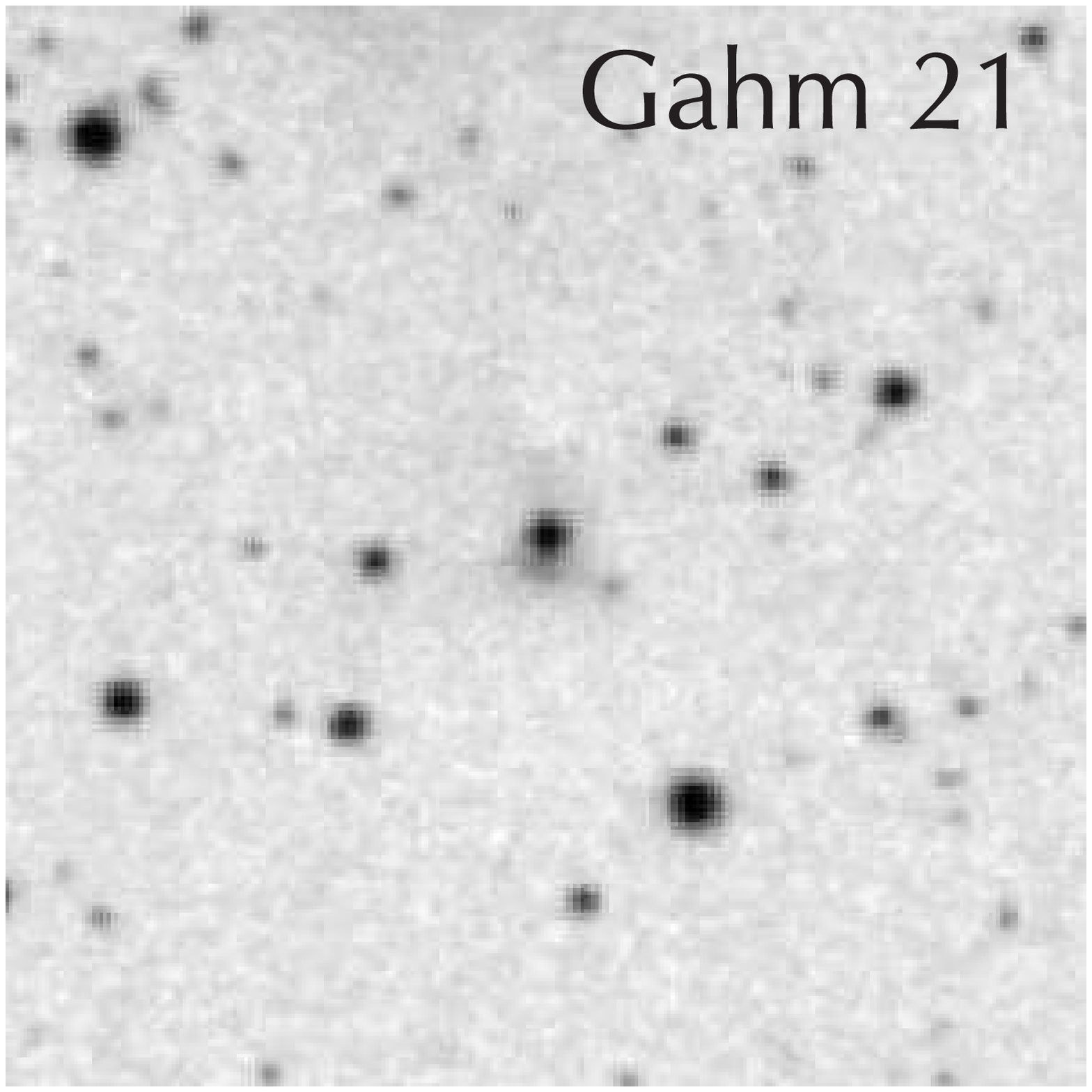,width=35mm,angle=0,clip=true}
}}
\vspace{-1mm}
\enlargethispage{2mm}
\captionb{1}{Identification charts.  The fields of
1.8\arcmin\,$\times$\,1.8\arcmin\ sizes are DSS2 red copies taken from
the Internet's Virtual Telescope SkyView.}
}

\newpage


\vbox{
\centerline{\psfig{figure=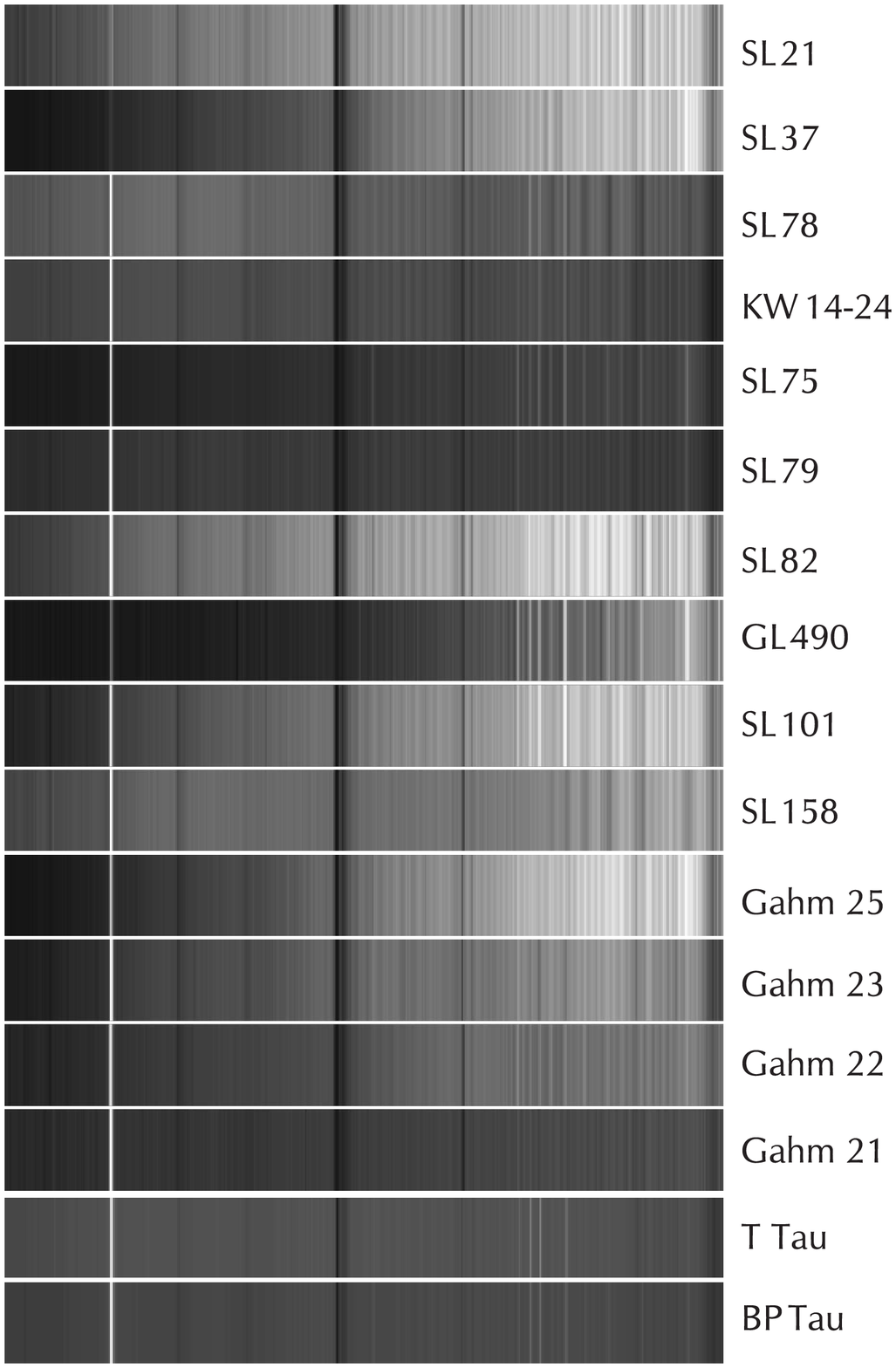,width=120mm,angle=0,clip=true}}
\vspace{0mm}
\captionb{2}{The widened spectra of the investigated stars. For
comparison, the spectra of two T Tauri type stars are given at the end.
The telluric H$_2$O and O$_2$ bands are not excluded.
}
}
\newpage

\vbox{\begin{center}
\psfig{figure=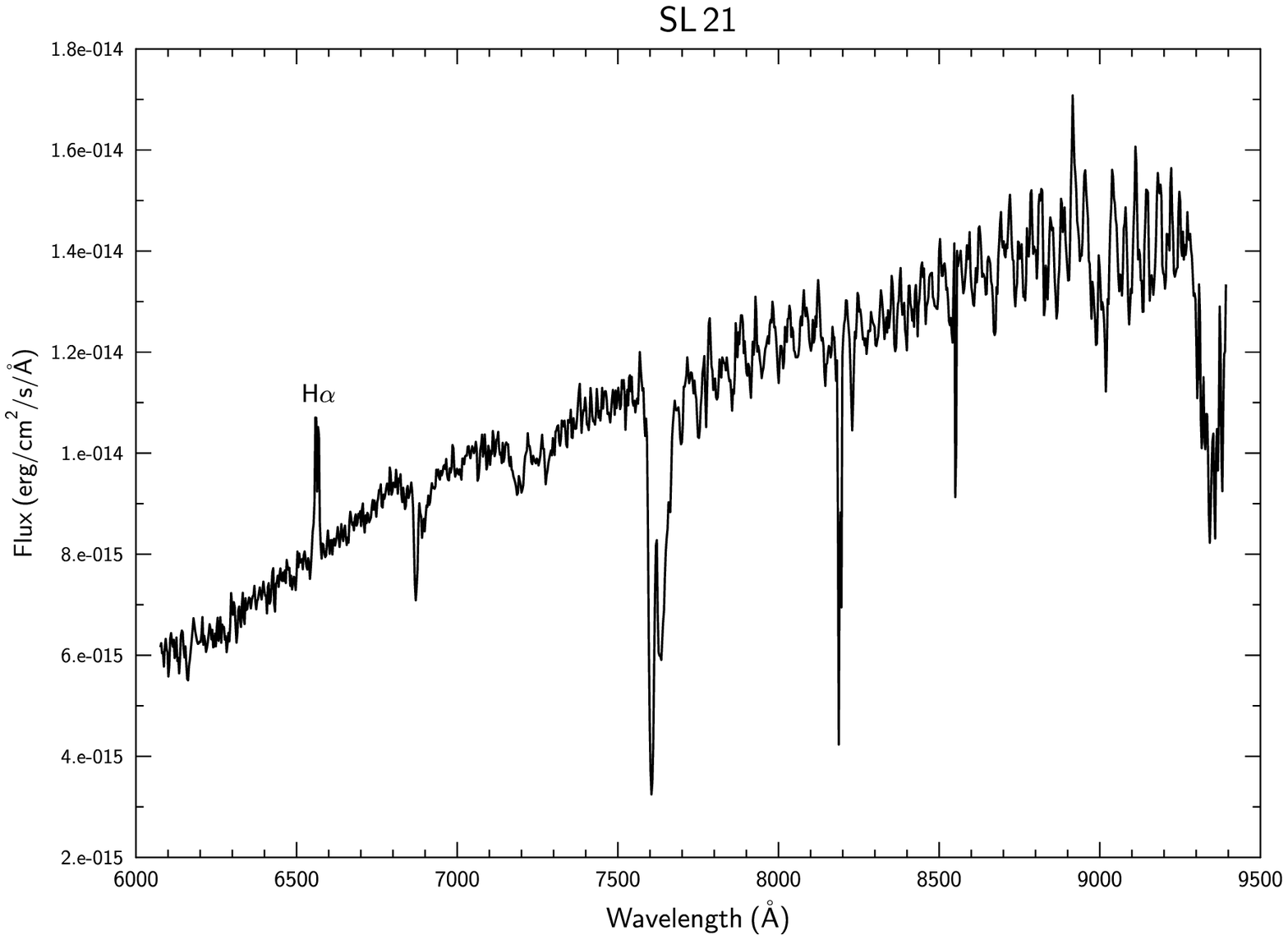,width=120mm,angle=0,clip=true}
\vskip2mm
\psfig{figure=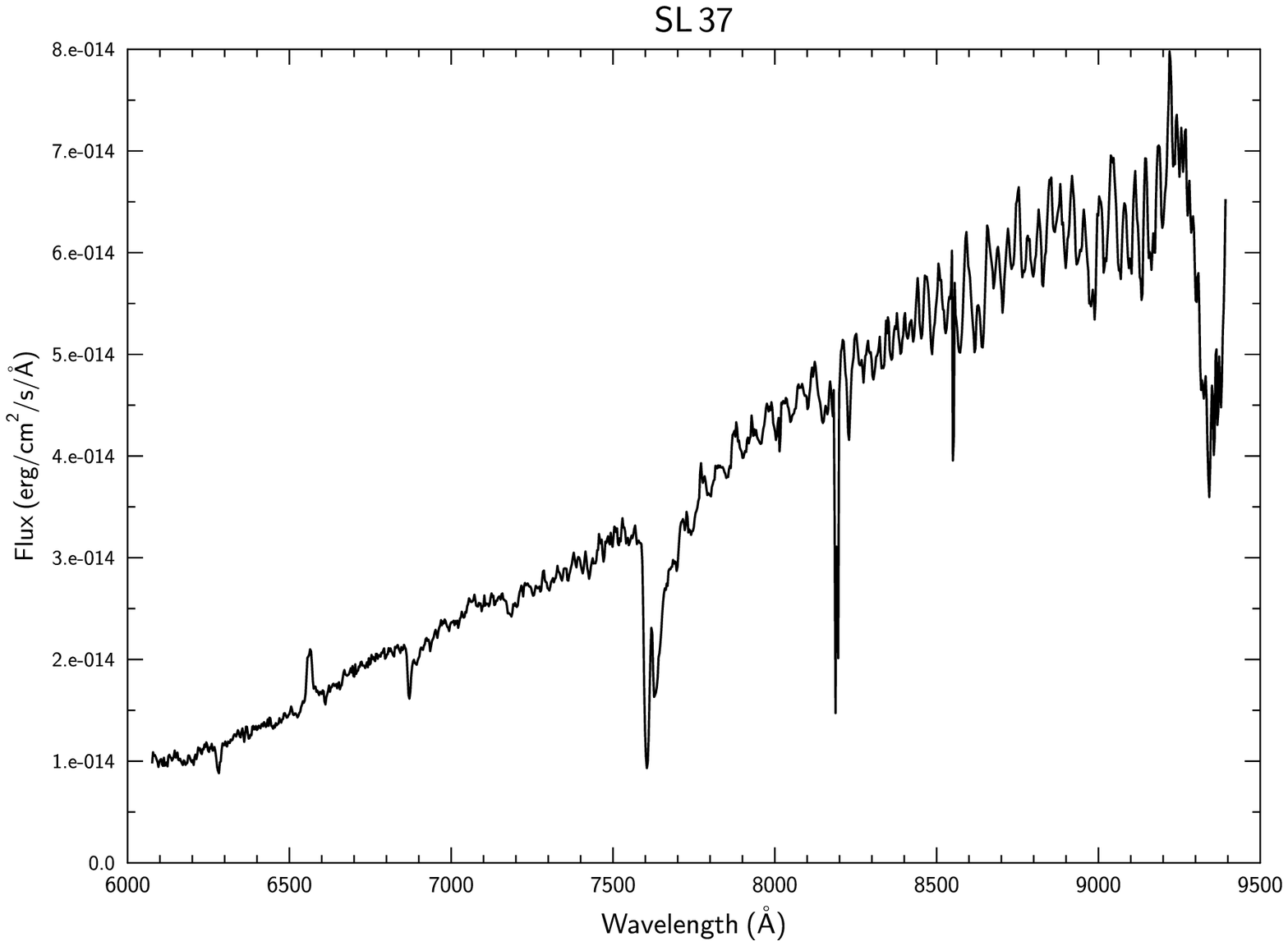,width=120mm,angle=0,clip=true}
\vskip.5mm
\captionc{3a and 3b}{Spectral energy distributions for the stars
SL\,21 and SL\,37.
}
\end{center}
}

\vbox{\begin{center}
{\psfig{figure=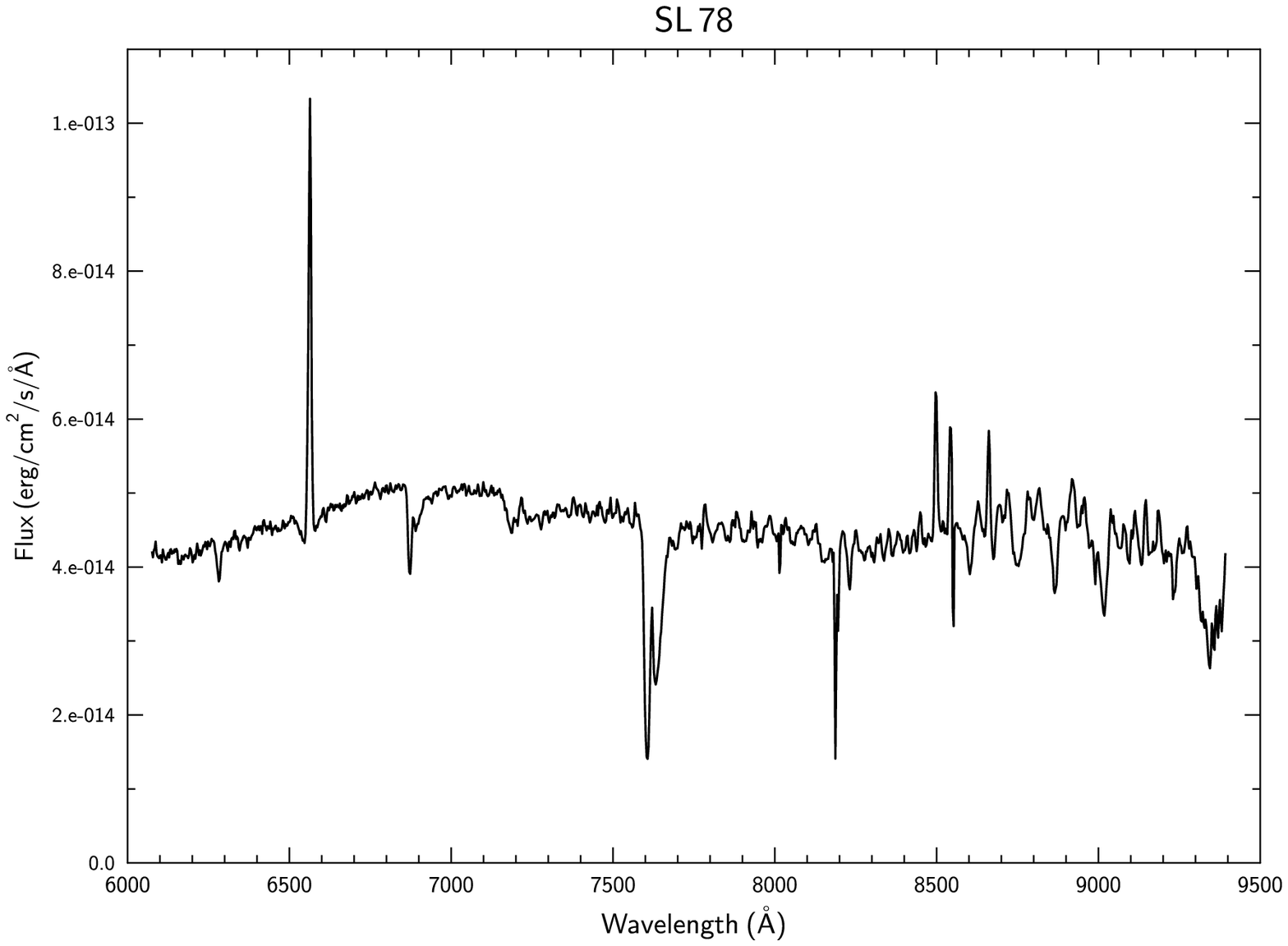,width=120mm,angle=0,clip=true}}
\vskip2mm
\psfig{figure=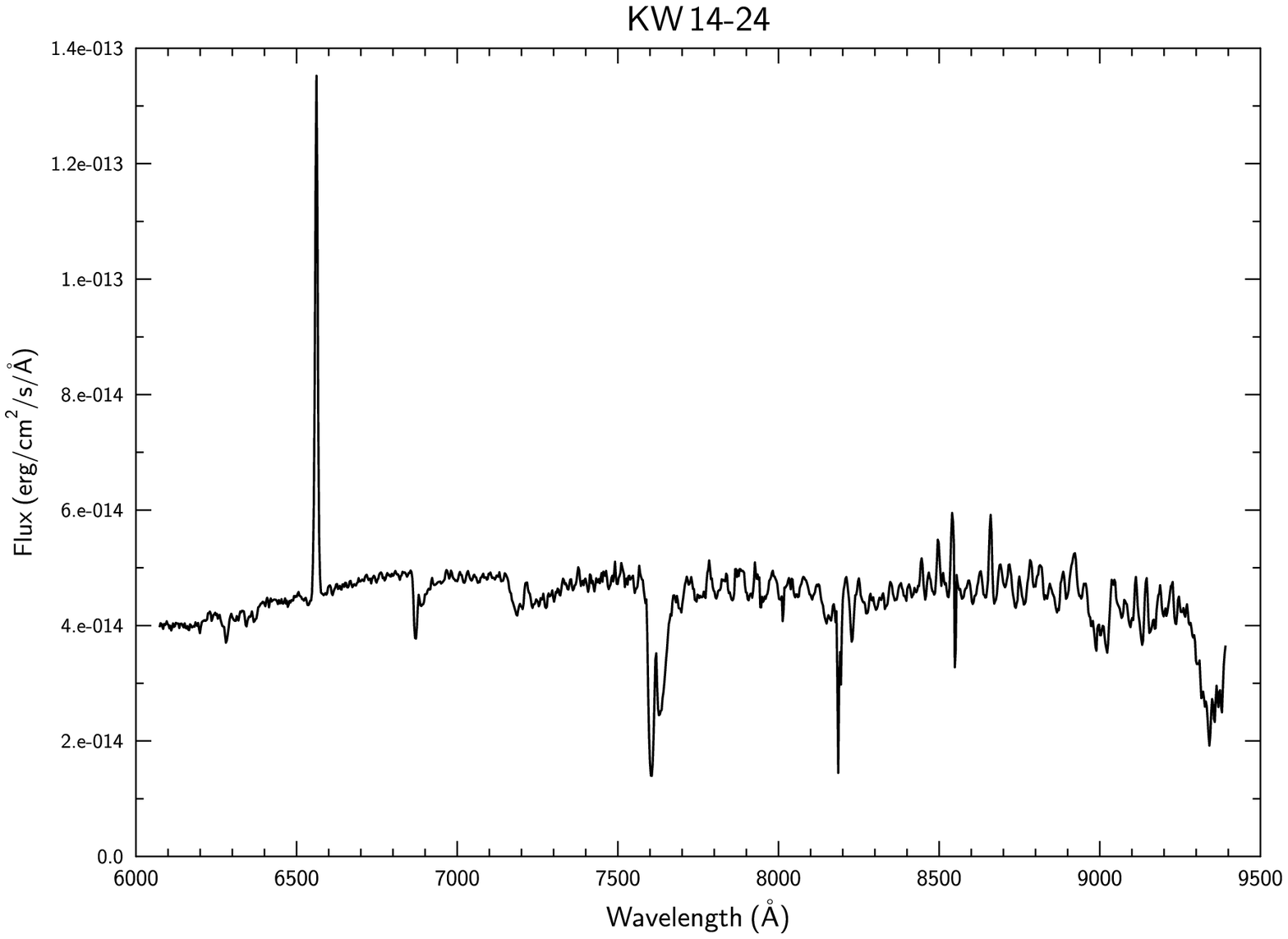,width=120mm,angle=0,clip=true}
\vskip.5mm
\captionc{3c~and~3d}{Spectral energy distributions for the stars
SL\,78 and KW\,14-24.}
\end{center}
}

\vbox{\begin{center}
{\psfig{figure=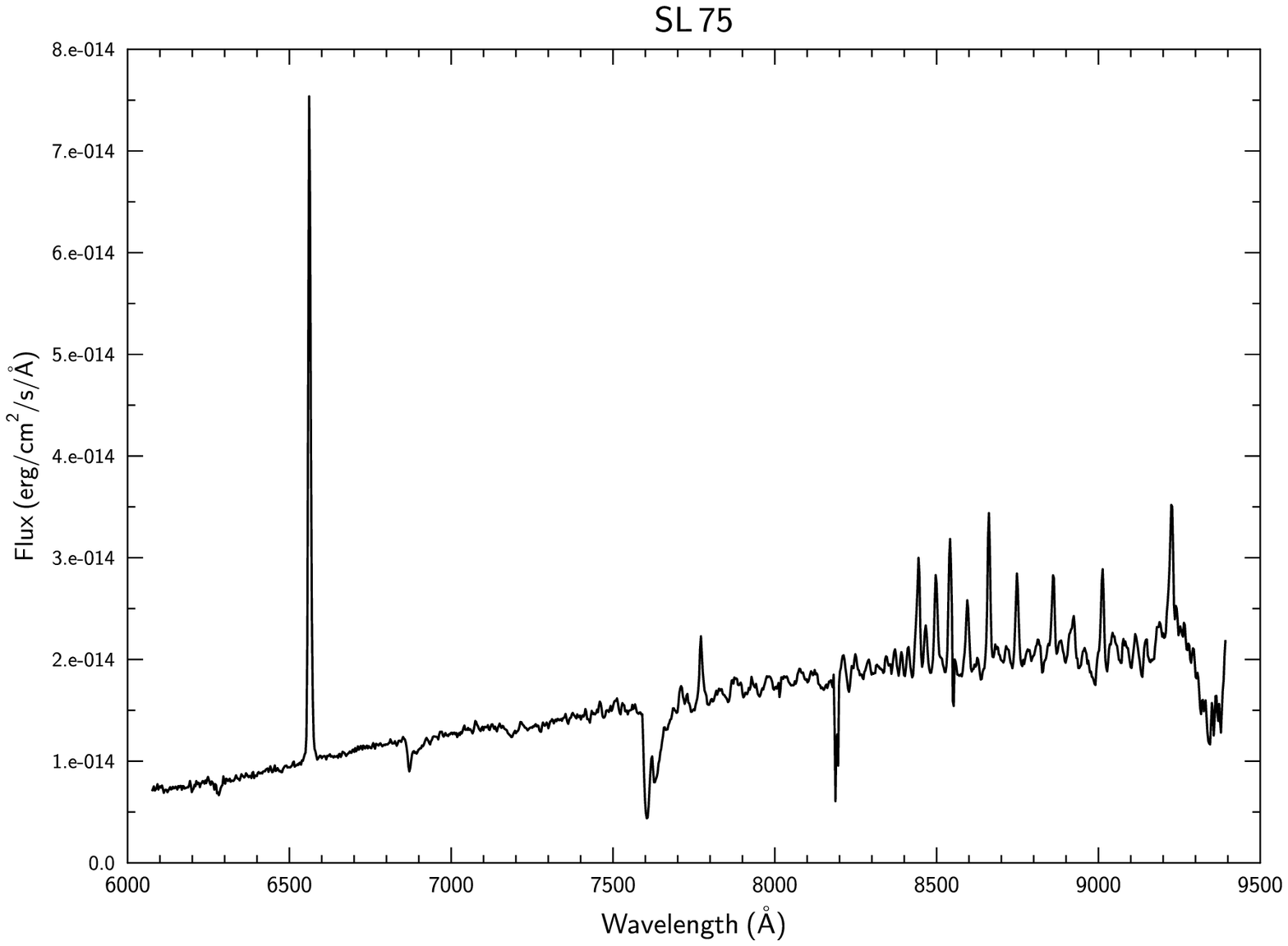,width=120mm,angle=0,clip=true}}
\vskip2mm
\psfig{figure=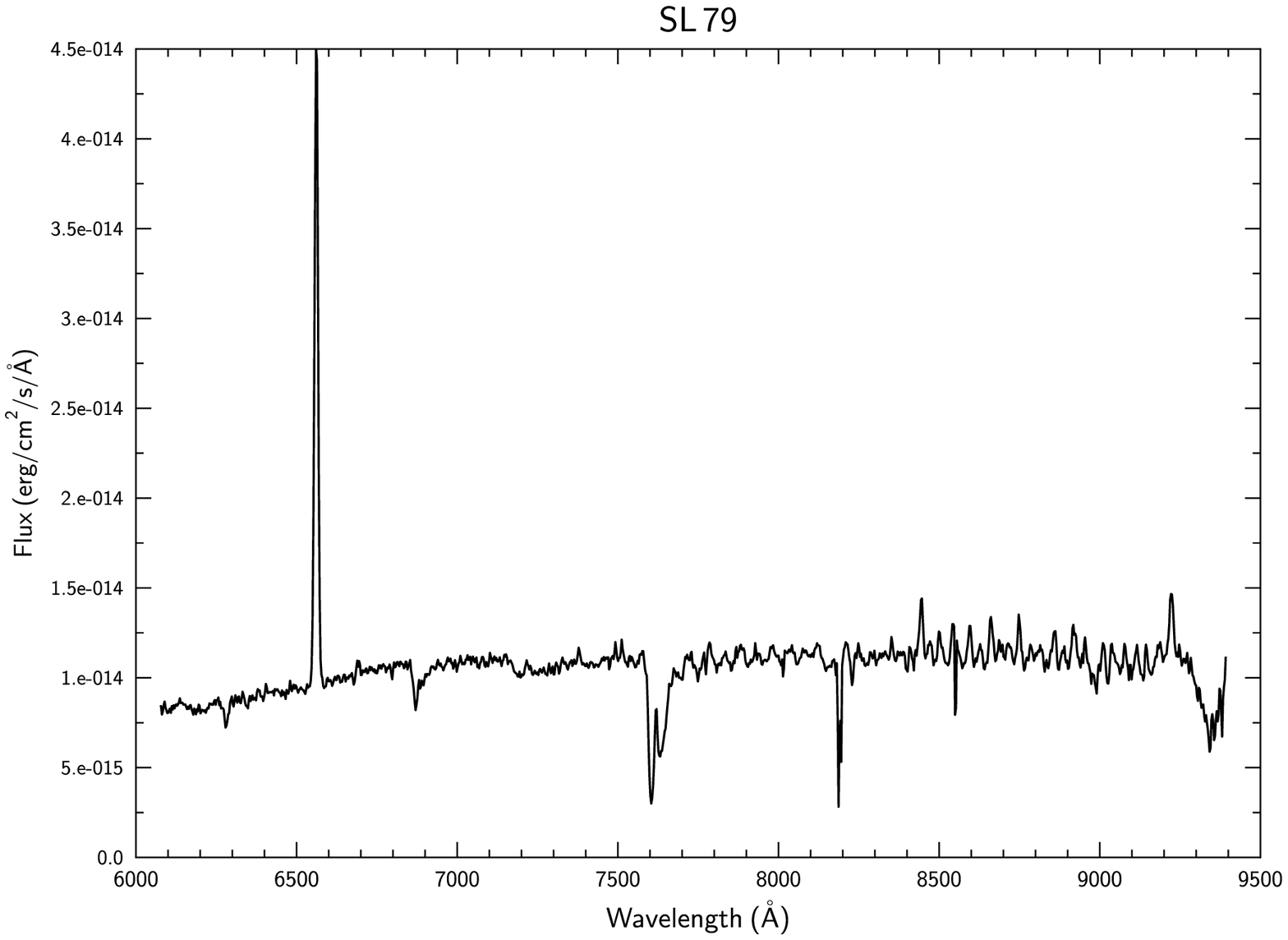,width=120mm,angle=0,clip=true}
\vskip.5mm
\captionc{3e~and~3f}{Spectral energy distributions for the stars
SL\,75 and SL\,79.}
\end{center}
}

\vbox{\begin{center}
{\psfig{figure=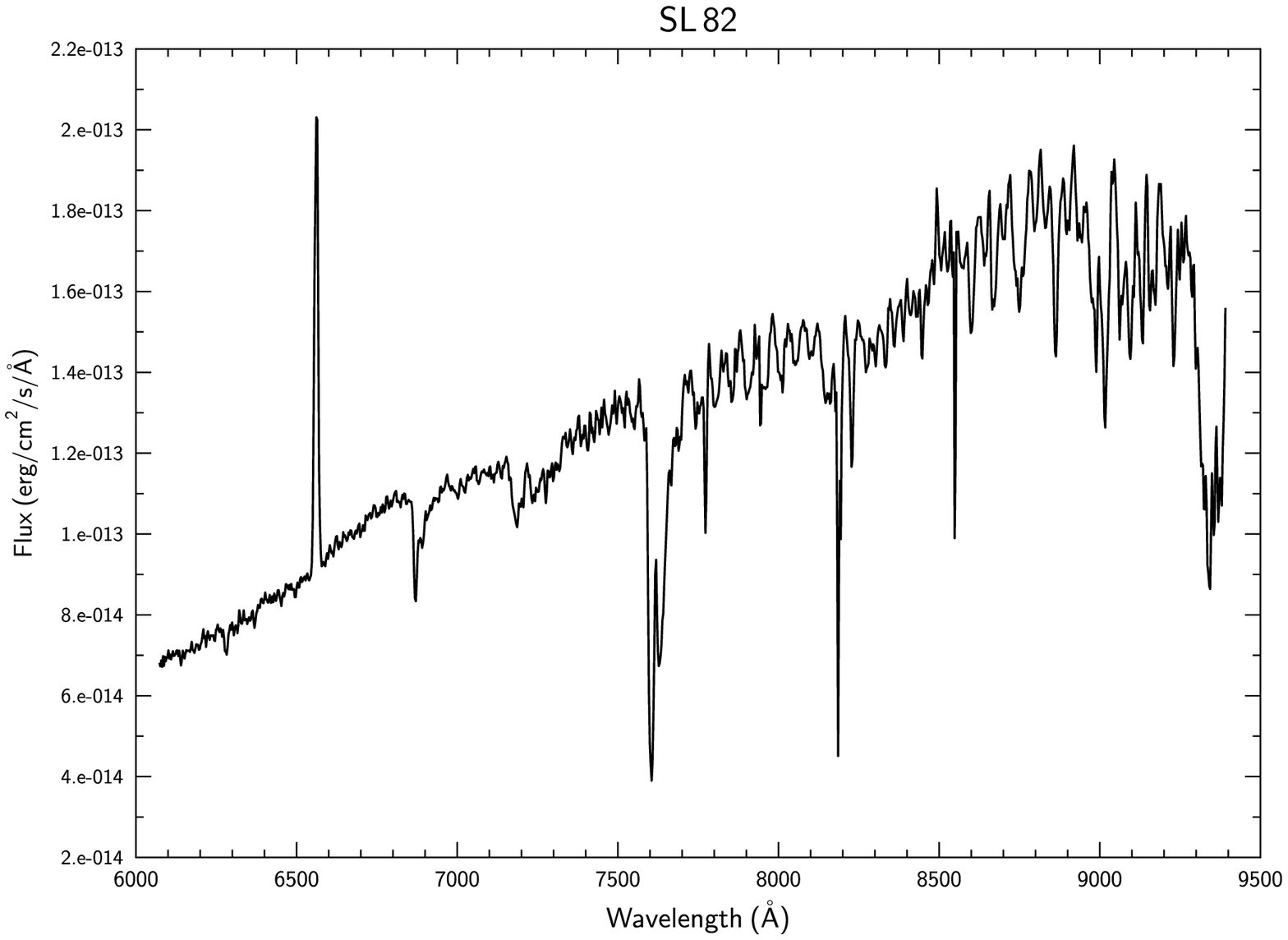,width=120mm,angle=0,clip=true}}
\vskip2mm
\psfig{figure=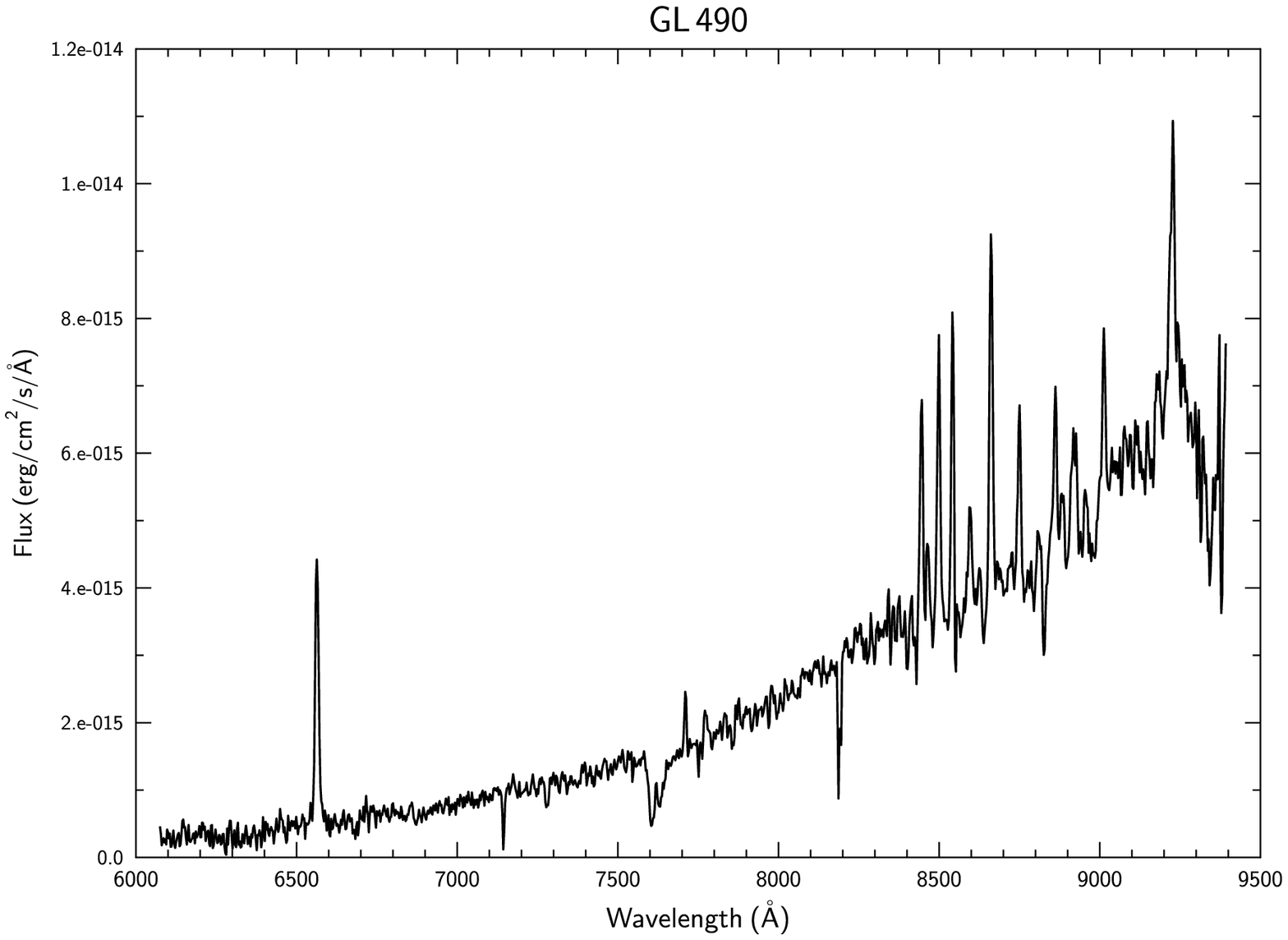,width=120mm,angle=0,clip=true}
\vskip.5mm
\captionc{3g~and~3h}{Spectral energy distributions for the stars
SL\,82 and GL\,490.}
\end{center}
}

\vbox{\begin{center}
{\psfig{figure=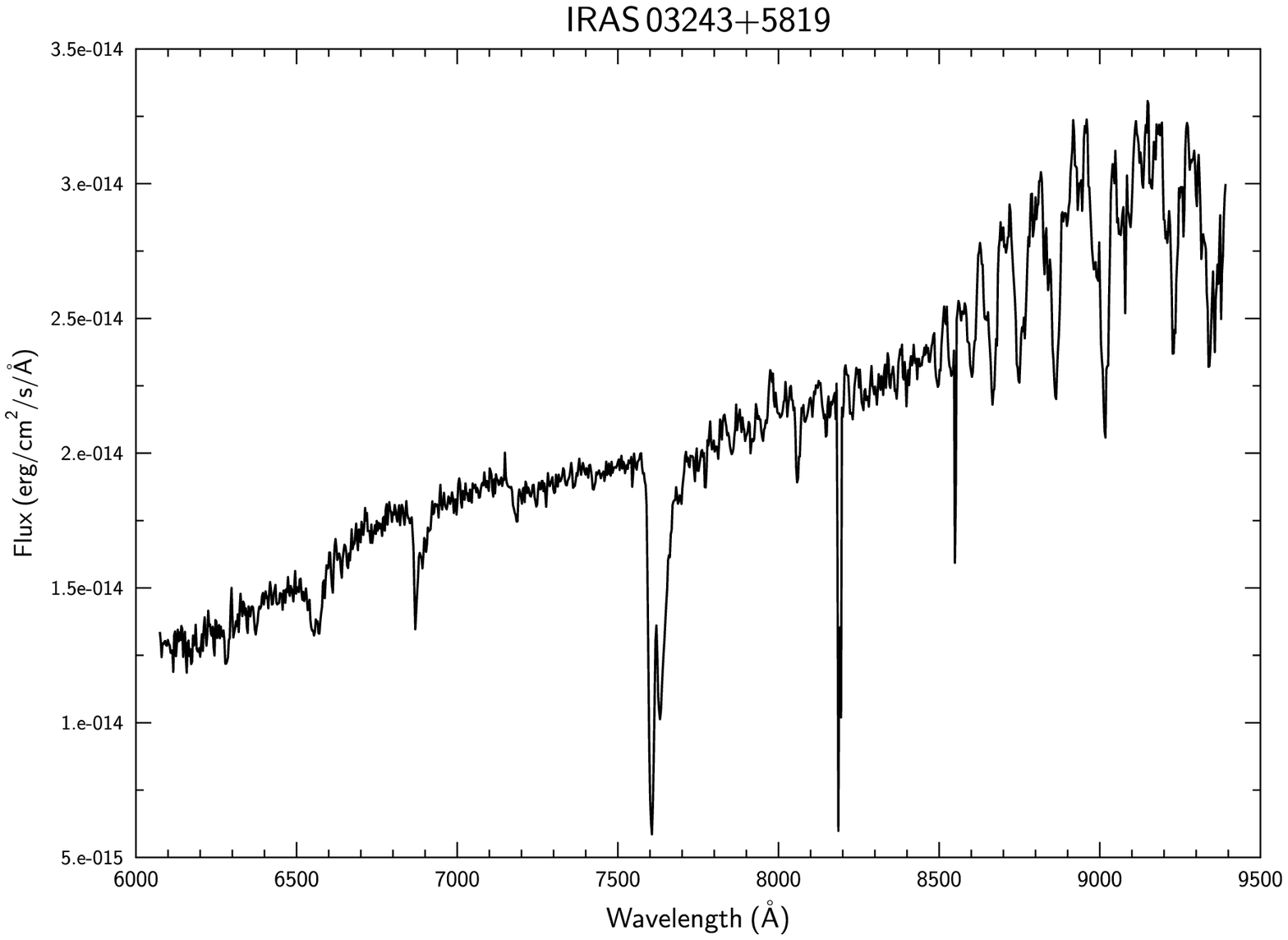,width=120mm,angle=0,clip=true}}
\vskip2mm
{\psfig{figure=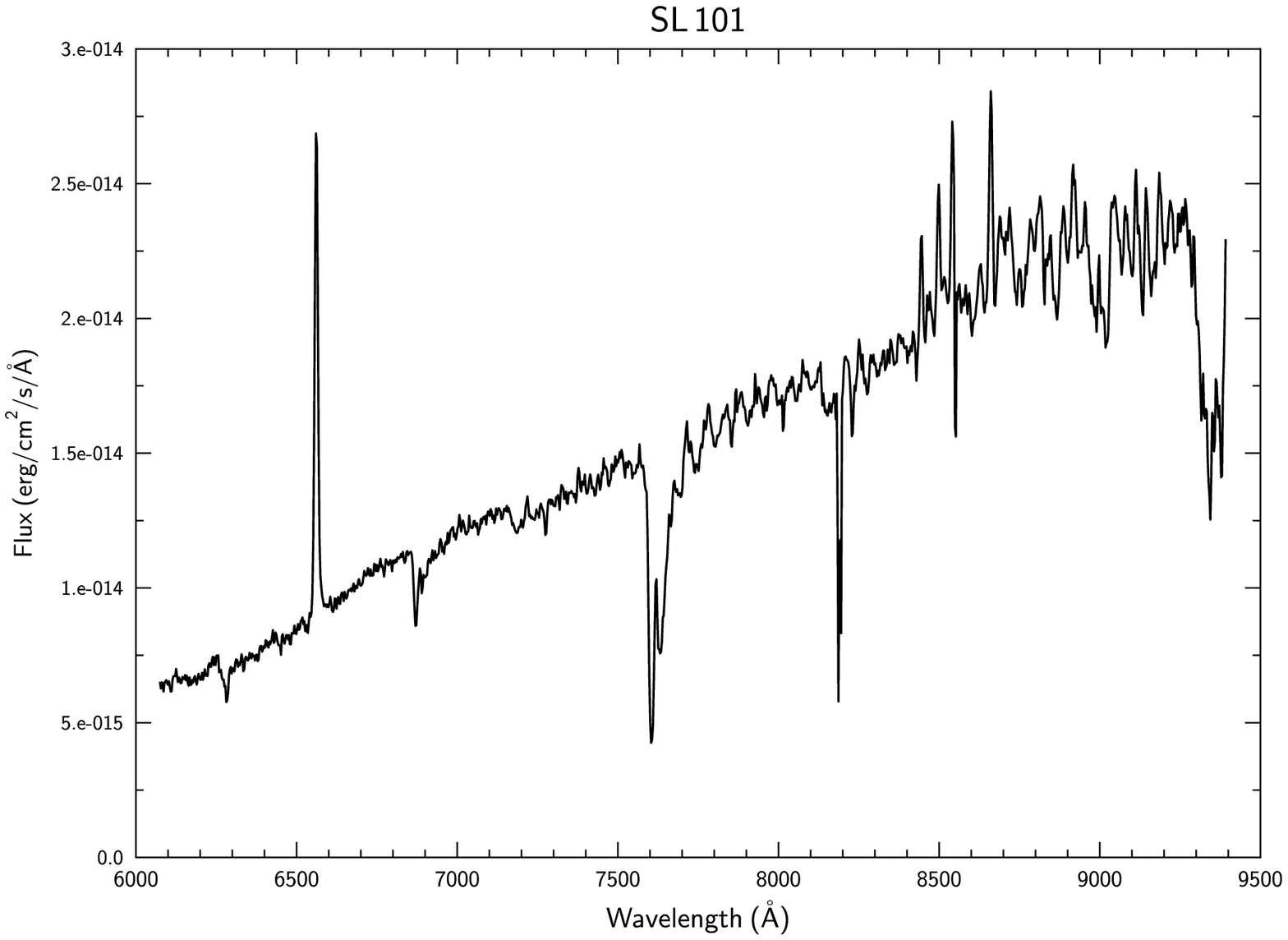,width=120mm,angle=0,clip=true}}
\vskip2mm
\captionc{3i~and~3j}{Spectral energy distributions for the stars
IRAS 03243+5819 and SL\,101.}
\end{center}
}

\vbox{\begin{center}
\psfig{figure=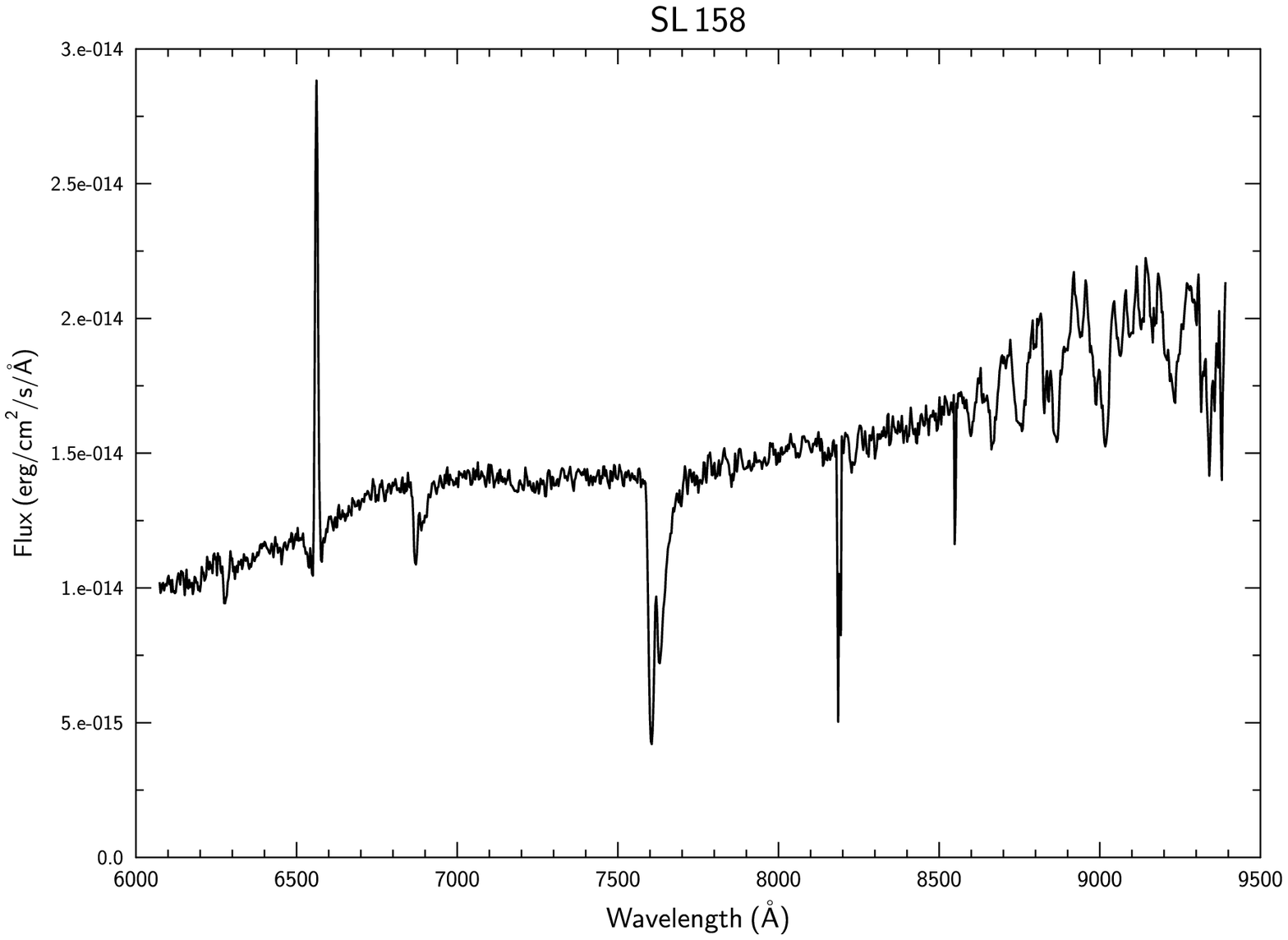,width=120mm,angle=0,clip=true}
\vskip.5mm
{\psfig{figure=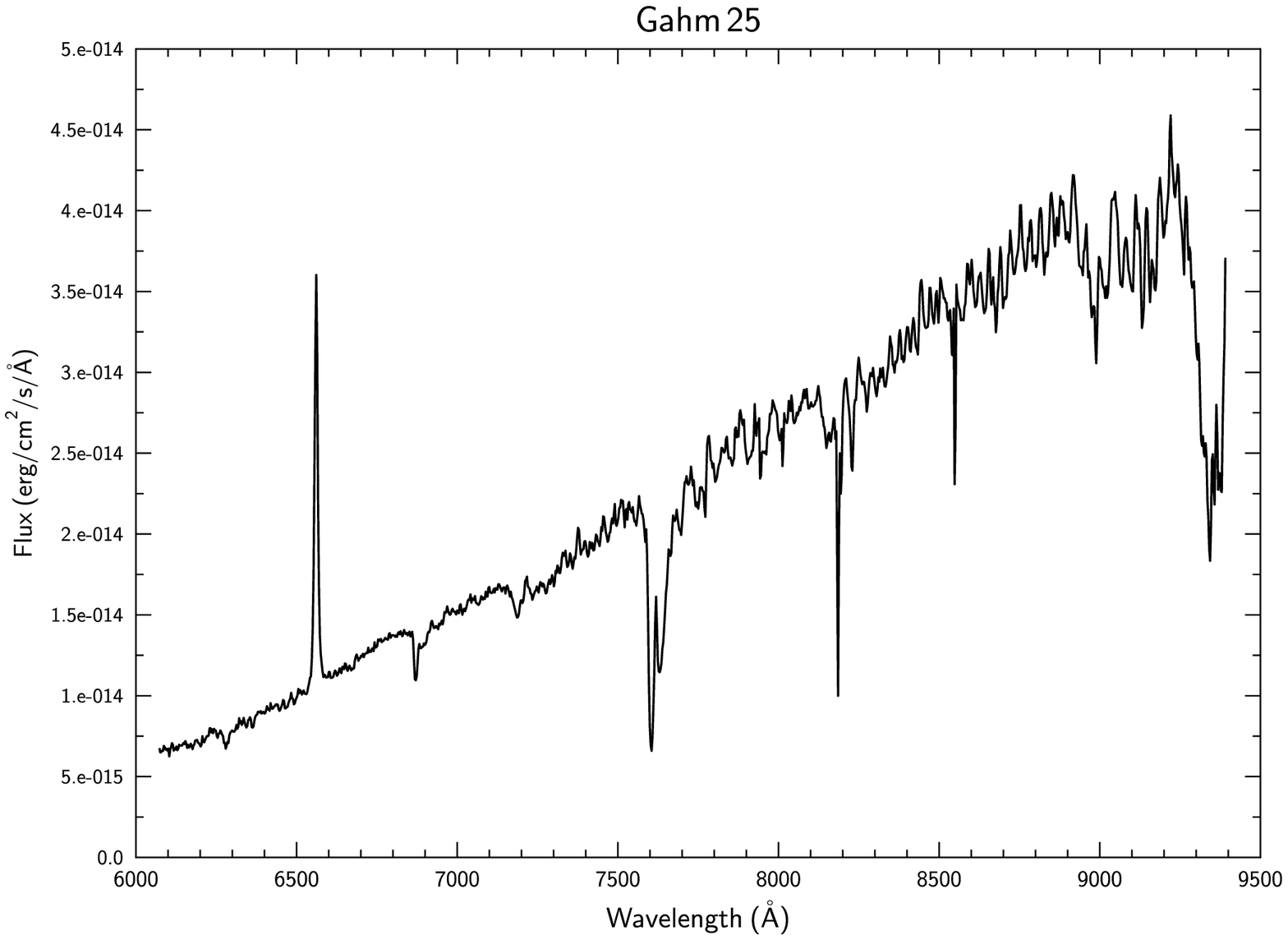,width=120mm,angle=0,clip=true}}
\vskip2mm
\captionc{3k~and~3l}{Spectral energy distributions for the stars
SL\,158 and Gahm 25.}
\end{center}
}

\vbox{\begin{center}
\psfig{figure=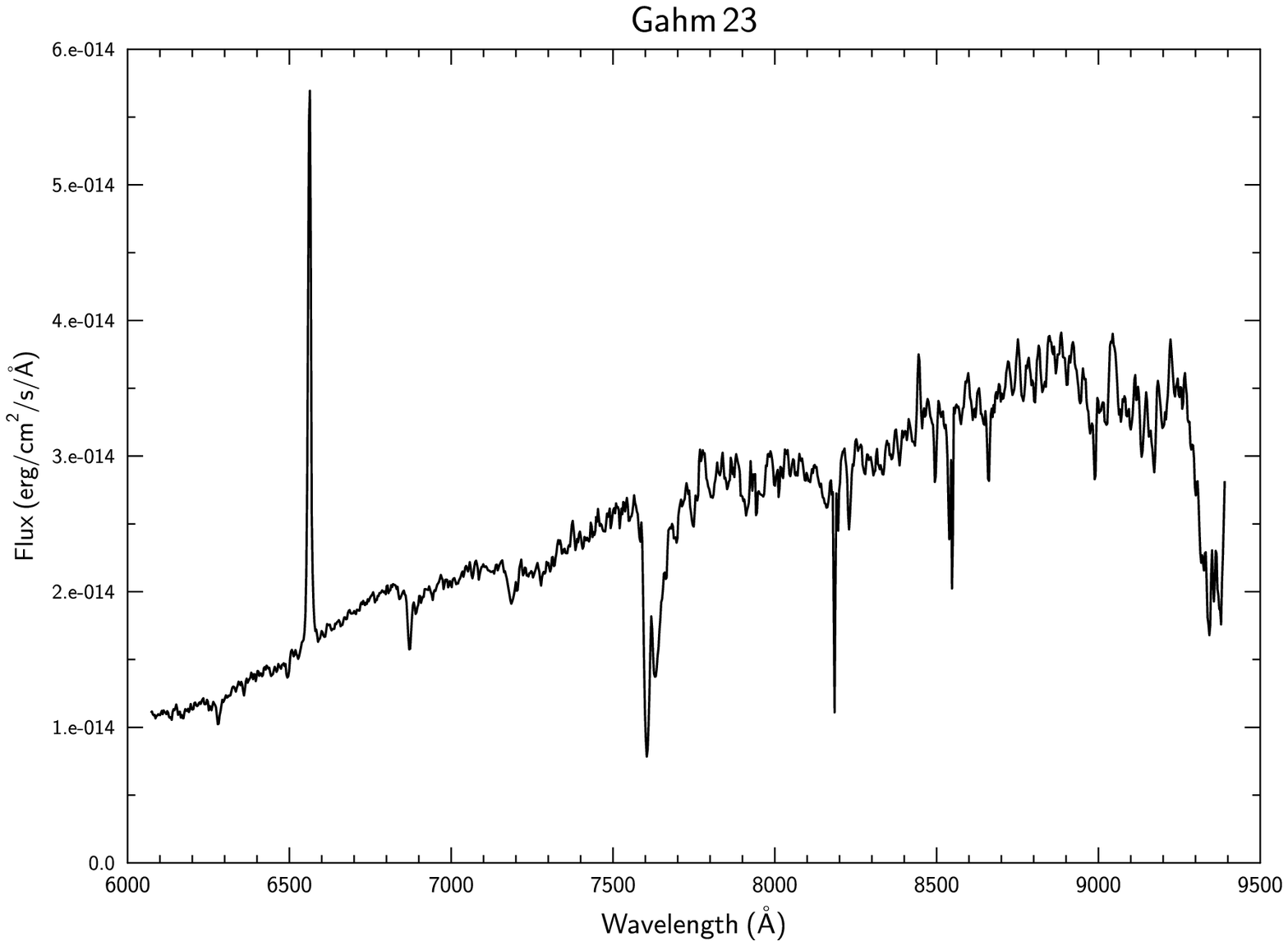,width=120mm,angle=0,clip=true}
\vskip.5mm
{\psfig{figure=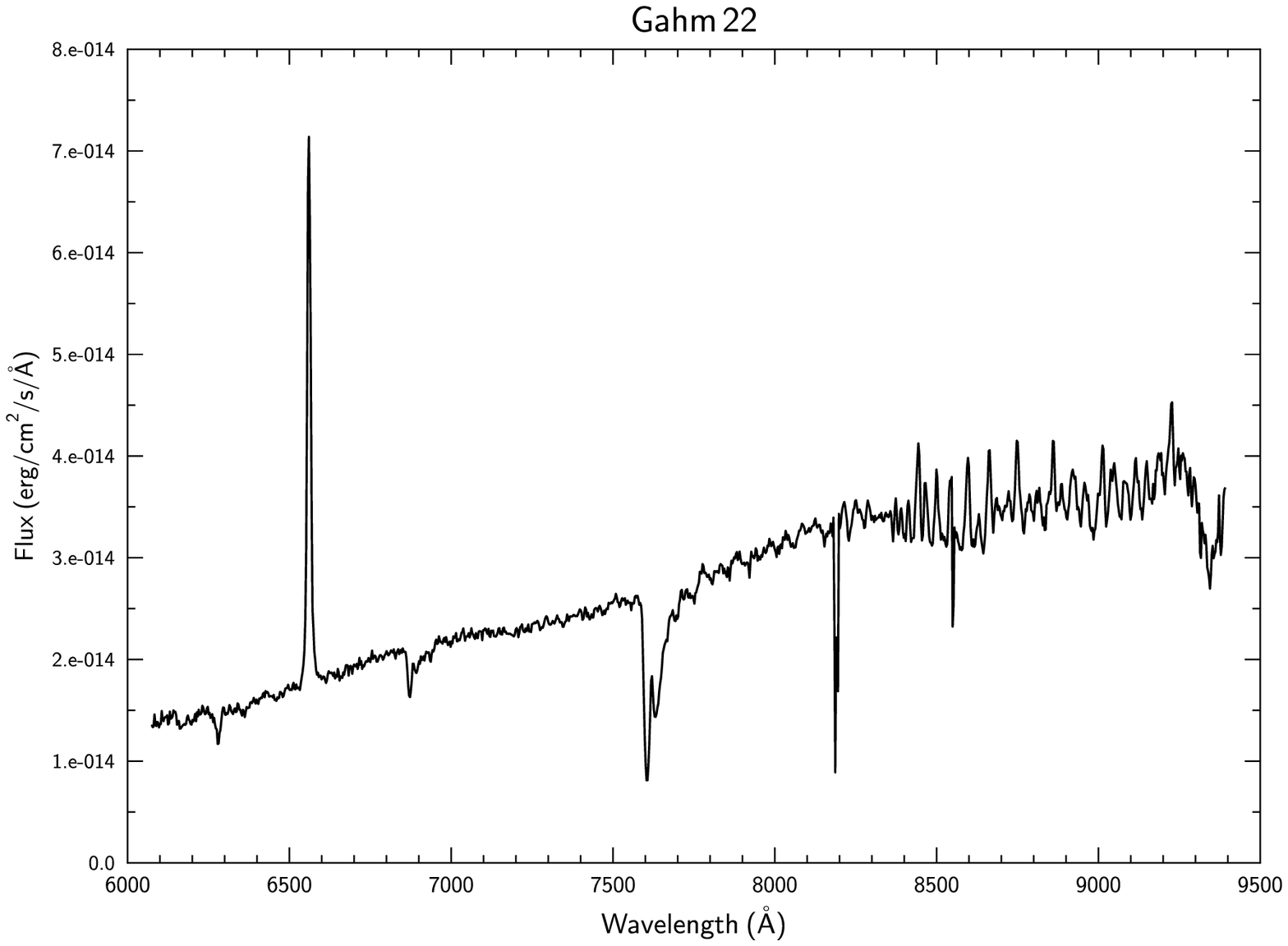,width=120mm,angle=0,clip=true}}
\vskip2mm
\captionc{3m~and~3n}{Spectral energy distributions for the stars
Gahm 23 and Gahm 22.}
\end{center}
}

\vbox{\begin{center}
\psfig{figure=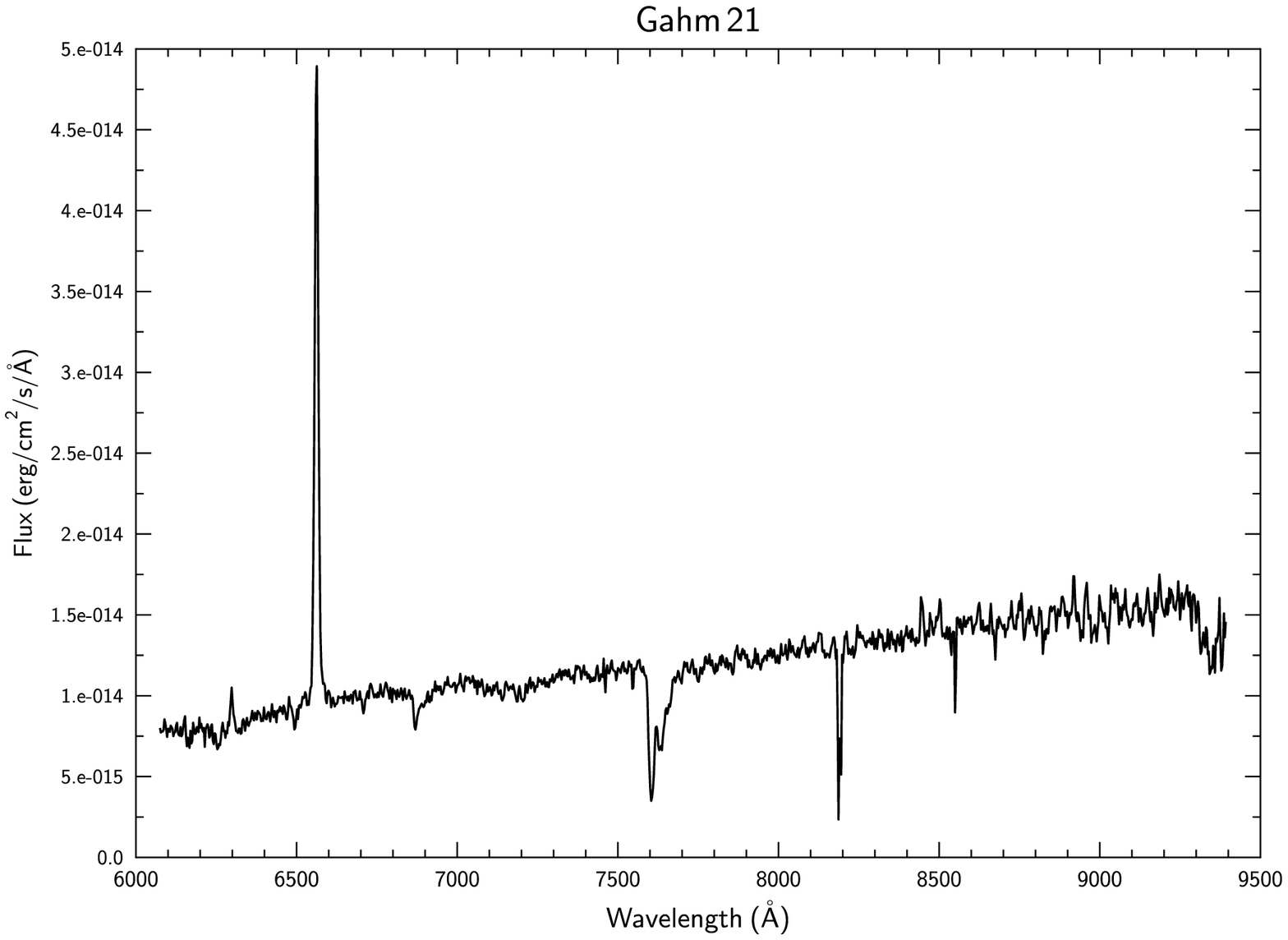,width=120mm,angle=0,clip=true}
\vskip.5mm
\psfig{figure=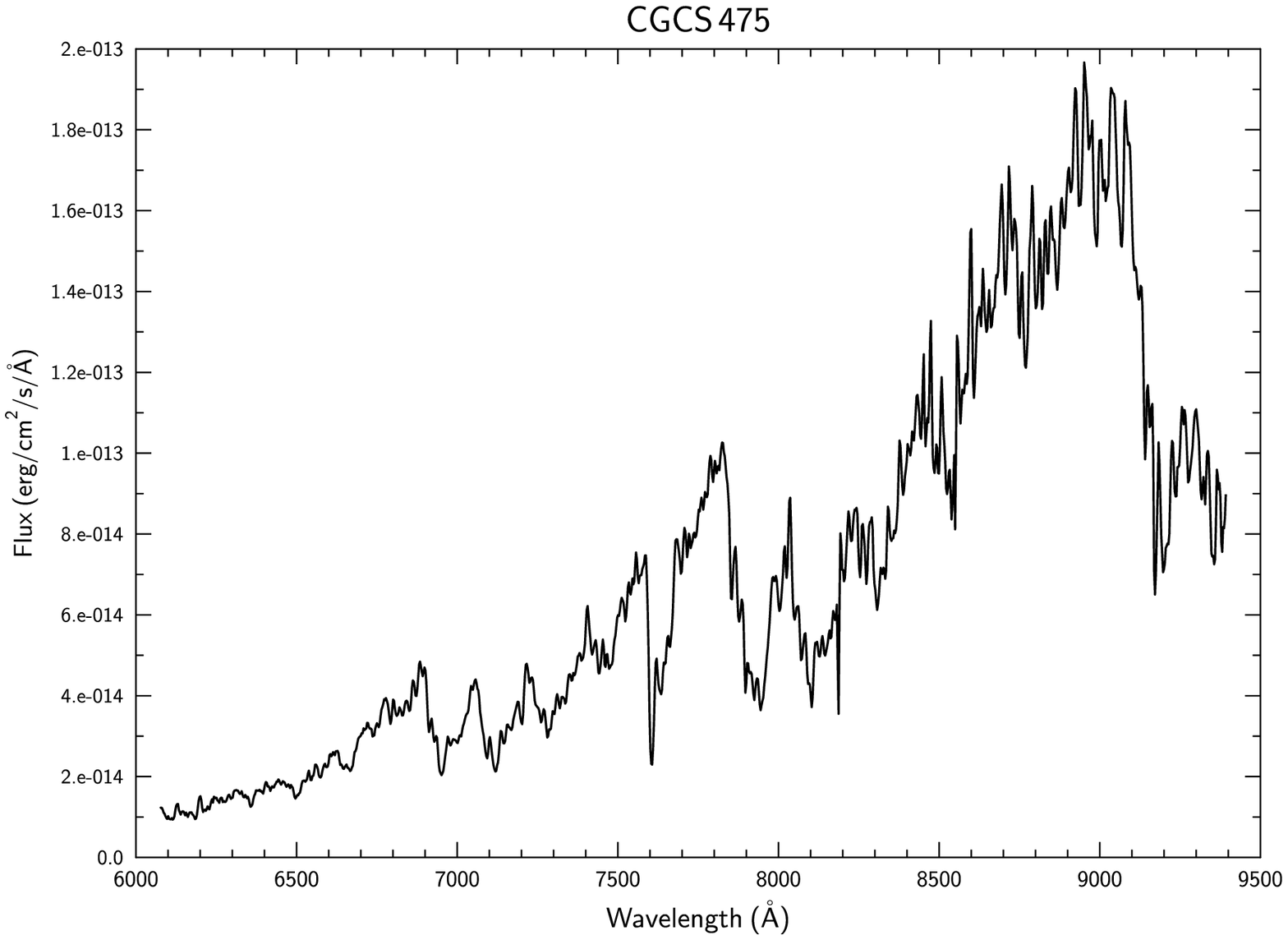,width=120mm,angle=0,clip=true}
\vskip.5mm
\captionc{3o~and~3p}{Spectral energy distributions for the stars
Gahm 21 and CGCS 475 (carbon star).}
\end{center}
}
\vskip3mm

\begin{table}[!th]
\begin{center}
\vbox{\footnotesize\tabcolsep=2pt
\parbox[c]{110mm}{\baselineskip=10pt
{\normbf\ \ Table 2.}{\norm\ Equivalent widths of emission lines and spectral
classification.}}
\begin{tabular}[t]{lrD..{-1}D..{-1}D..{-1}D..{-1}D..{-1}D..{-1}l}
\tablerule
\multicolumn{1}{c}{Star}  &
\multicolumn{1}{c}{Obs. date} &
\multicolumn{1}{c}{EW\,6563} &
\multicolumn{1}{c}{EW\,8446} &
\multicolumn{1}{c}{EW\,8498} &
\multicolumn{1}{c}{EW\,8542} &
\multicolumn{1}{c}{EW\,8662}&
\multicolumn{1}{c}{EW\,9226} &
\multicolumn{1}{c}{Spectral}  \\
      &  & \multicolumn{1}{c}{H$\alpha$} &
\multicolumn{1}{c}{O\,I} &
\multicolumn{1}{c}{Ca\,II} &
\multicolumn{1}{c}{Ca\,II} &
\multicolumn{1}{c}{Ca\,II} &
\multicolumn{1}{c}{P9} &
\multicolumn{1}{c}{class} \\
\tablerule
SL\,21    & 2007-10-23 & -4.6  &       &       &       &       &         & G0e\,(0.2) \\
SL\,37    & 2007-10-23 & -6.5  &       &       &       &       &  -1.6   & A5e\,(0.2) \\[1pt]
SL\,78    & 2007-10-23 & -10.2 & -1.7 & -3.4 & -1.8 & -2.2 &         & F5e\,(0.5) \\[1pt]
KW\,14-24 & 2007-10-21 & -20.6 & -0.8 & -1.4 & -2.2 & -2.1 &         & G5e\,(1)   \\[1pt]
SL\,75    & 2007-10-23 & -64.8 & -4.8 & -4.1 & -4.5 & -6.7 & -12.2  & G0e\,(1.5) \\[1pt]
SL\,79    & 2007-10-23 & -50.6 & -3.0 & -1.1 & -0.4 & -2.2 & -3.9   & G0e\,(1)   \\[1pt]
SL\,82    & 2007-10-21 & -17.2 &       & -0.8 &       &       &         & G5e\,(0.5) \\[1pt]
GL\,490    & 2007-10-22 & -86.2 & -8.2 & -10.6& -11.1& -14.7& -9.6   & Ae\,(2)    \\[1pt]
IRAS\,$^*$  & 2007-10-22 &    &      &      &      &      &        & Ae\,$^*$ \\[1pt]
SL\,101   & 2007-10-23 & -26.0 & -1.5 & -2.0 & -2.0 & -4.0 &         & G0e\,(0.5) \\[1pt]
SL\,158   & 2007-10-22 & -11.8 &       &       &       &       &         & Ae\,(0.5)  \\[1pt]
Gahm 25   & 2007-10-21 & -29.5 &       &       &       &       & -0.5   & K2e\,(0.5) \\[1pt]
Gahm 23   & 2007-10-21 & -28.4 & -1.3 &       &       &       & -0.8   & Ge\,(1)    \\[1pt]
Gahm 22   & 2007-10-22 & -42.2 & -1.6 & -1.4 & -1.8 & -0.8 & -1.6   & Ae\,(1.5)  \\[1pt]
Gahm 21   & 2007-10-22 & -59.0 & -1.1 & -1.0 &       &       &         & Ke\,(1.5)  \\
\noalign{\vskip1mm}
T Tau     & 2007-10-21 & -91.8 & -4.0 & -13.9& -10.8& -14.0& -1.9   &  K1e \\[1pt]
BP Tau    & 2007-10-21 & -149.1& -3.1 & -9.3 & -9.0 & -11.7& -3.9   &  K7e  \\[1pt]
CY Tau    & 2007-10-21 & -36.0 &       &       &       &       &         &  M1e \\[1pt]
DI Tau    & 2007-10-22 &    0.4 &       &       &       &       &         &  M0e \\
\tablerule
\end{tabular}
}
\end{center}

\noindent $^*$~~{\footnotesize IRAS 03243+5819, emission core in the center
of H$\alpha$ absorption line.}

\end{table}

\vskip1mm

The equivalent widths of the prominent emission features (H$\alpha$,
Ca\,II triplet, O\,I at 8498 \AA\ and P9 at 9226 \AA) were measured with
the IRAF `splot' utility.  They are taken across the whole line,
including any absorption wings, so positive EWs will result when there
is emission in just the line cores.  The mean values of three measures
are given in Table 2. EWs were measured also in the spectra of four T
Tauri stars, T Tau, BP Tau, CY Tau and DI Tau, having very different
emission-line intensities.  During our observations the star DI Tau did
not show any emission, even in H$\alpha$.

\sectionb{3}{DISCUSSION ON INDIVIDUAL OBJECTS}

In this section we give more information about the investigated
objects and discuss the results of spectral classification and emission
line intensities.

\vskip1mm

{\bf SL\,21 = 2MASS J02231856+6125416}

The object is located in the dark cloud T\,879
to W from the H\,II region W3/4.  In Paper II, according to the LSR
radial velocities of the surrounding CO clouds, the cloud is attributed
to the Perseus spiral arm.  However, in this direction the lower
velocity components of some clouds are also prominent, so we do not
exclude that SL\,21 is an object of the Local arm.  The inspection of
its images in SkyView shows that the object is double.  The components
of similar brightness are separated by $\sim$\,1.5\arcsec\ and are
located along the NE--SW direction.  In the $B$ passband the NE object
is slightly fainter, in $R$ and $I$ both objects are of equal
brightness, and in $J$, $H$ and $K$ the NE object dominates.  This means
that this object is the cooler component of the system.  If the object
is in the Local arm at a distance of 1 kpc (the distance of the Cam OB2
association), then both components are separated by 1500 AU.  An
infrared object IRAS 02198+6111 (Kerton \& Brunt 2003), located within
about 3$\arcmin$, has no relation to SL\,21, it corresponds to another
2MASS object.  In our spectrum both components are not separated.  It
exhibits a noticeable H$\alpha$ emission, the lines of O\,I and Ca\,II
are not seen.  Probably, one or both components are T Tauri or post T
Tauri stars of early G class.

\vskip1mm

{\bf SL\,37 = 2MASS J02271602+6200506}

The object is located within the same cloud T\,879, in the outskirts
of the H\,II region W3.  First recognized as an H$\alpha$ emission
object in the objective-prism spectra by Dolidze (1975, her number
02-075).  The star was included in the emission-star catalog of
Kohoutek \& Wehmeyer (1997, star number 11-56).  Faint X-ray source
(Rosat).  Our classification gives A5e(0.2) spectral type.  Probably, it
is a Herbig Ae/Be star related to clustering of stars in the H\,II
region IC\,1795, belonging to the Perseus arm.

\vskip1mm

{\bf SL\,78 = 2MASS J02514696+5542014}

The object is located within a dark cloud T\,931 belonging to the Cam
OB1 association dust layer of the Local arm.  No object is present in
the Simbad database within 2\arcmin\ of its position.  According to our
spectra, the object exhibits quite strong emissions in H$\alpha$, O\,I
and Ca\,II.  Other criteria prove it to be of spectral class F5,
consequently this is a star of intermediate temperature between
classical T Tauri stars and Herbig Ae/Be stars.  The star is present
in the MSX catalog (G139.3681-03.2822) with a flux at 8.3 $\mu$m of
0.192 Jy (quality 3). The energy distribution curve of the star is
similar to YSOs of Class III (see Paper II).

\vskip1mm

{\bf KW 14-24 = 2MASS J03012159+6028566}

An entry in the emission-star catalog of Kohoutek \& Weh\-meyer (1997).
First discovered by Dolidze (1975, her number 02-155).  Recently the
H$\alpha$ emission was confirmed by the IPHAS photometric survey (Witham
et al. 2008).  Faint X-ray source (Rosat).  The star is located at a
bright rim of the ionized gas at the left edge of the H\,II region W5
belonging to the Perseus arm (Ogura et al. 2002; Karr \& Martin 2003b).
It may be one of the objects formed by a triggered star formation
mechanism due to compression of gas and dust by ionization fronts.  In
our spectrum the emission line H$\alpha$ is very strong, and the O\,I
and Ca\,II emissions are much fainter, spectral class G5.

\vskip1mm

{\bf SL\,75 = 2MASS J03032586+6023095}

No object is present in the Simbad database within 2\arcmin\ of its
position.  Thus, this is a new YSO in one of the clumps of the dust
cloud T\,912 at the left edge of the H\,II region W5.  A broad
15\arcsec\ long protrusion or jet is seen north of the object.  Probably
this is a classical T Tauri star of spectral class at G0 with strong
emissions in H$\alpha$, O\,I, Ca\,II and P9 lines, similar in strengths
to the emissions in the prototype star T Tau.

\vskip1mm

{\bf SL\,79 = 2MASS J03104626+5930035}

This object is located in a faint emission nebula LBN 140.07+1.64 (Lynds
1965), near the southern edge of \hbox{Sh 2-202} belonging to the Local
arm.  However, this LBN nebula shows the velocity of the Perseus arm.
The counterpart of the object is IRAS 03068+5918 (Karr \& Martin 2003a;
Kerton \& Brunt 2003).  Nearby, at an angular distance of about
8\arcsec\ in the NE direction another very similar object, 2MASS
J03104707+5930082, is located.  This object in the $V$ passband is
fainter by 1.3 mag, but with increasing $\lambda$ the difference
diminishes, and in $J$ and $H$ both sources are of equal intensity.  In
the $K_s$ passband the NE object is again fainter by 0.8 mag.  Our
spectra correspond to the right, brighter object.  The star seems to be
a typical T Tauri star of G-type with very strong H$\alpha$ emission
line and fainter emissions in O\,I, Ca\,II and P9.

\vskip1mm

{\bf SL\,82 = 2MASS J03172590+6009417}

The star is also known as IRAS 03134+5958 and CPM\,7 (Campbell et al.
1989).  It was first described as an emission-line object and included
in the catalog of emission-line stars of the Orion population by
Herbig \& Bell (1988).  It is located at the center of the Sh\,2-202
nebula, near the questionable open cluster Stock 23.  In the Palomar
atlas blue and red images it looks like a comet-like object surrounded
by a nebulous coma with a diameter of $\sim$\,0.5\arcmin, with a broad
short tail directed to NE.  Our spectra confirm its emission-line status
with a moderately strong emission in H$\alpha$ and faint emission in
Ca\,II, spectral class is $\sim$\,G5.

\vskip1mm

{\bf GL\,490 = 2MASS J03273876+5847000}

A heavily reddened ($A_V$\,$\approx$\,35 mag, Alonso-Costa \& Kwan 1989)
high-mass young stellar object (IRAS 03236 +5836), embedded in the
densest part of the dust cloud T\,942.  GL\,490 was discovered in the
US Air Force Geophysics Laboratory (AFGL) Infrared Sky Survey using 16.5
cm telescopes flown above the atmosphere on rocket probes.  The Price \&
Walker (1976) catalog presents fluxes of GL\,490 in the 11 and 20 $\mu$m
passbands.  The star belongs to the Cam OB1 association layer of the
Local arm.  It is a YSO of Class I:  surrouded by circumstellar gas and
dust shell which gives a strong far-infrared excess, see its spectral
energy distribution in Paper II.  The spectrum in the interval 600--1000
nm was observed by McGregor et al.  (1984):  H$\alpha$, Paschen series,
O\,I, Ca\,II triplet and [S\,III] lines are in emission.  The lines
Br$\alpha$, Br$\gamma$ and Pf$\gamma$ are also in emission (see Simon et
al. 1983; Bunn et al. 1995 and references therein).  Our spectrum shows
strong emissions in H$\alpha$, O\,I, Ca\,II and P9.

\vskip1mm

{\bf IRAS 03243+5819 = 2MASS J03281460+5829374}

This star is located in the dust cloud T\,942, only $\sim$\,20\arcmin\
from GL\,490.  In Papers II and III it was not included in the lists of
suspected YSOs since in the $J$--$H$ vs.  $H$--$K_s$ diagram the star
was found to be located below the intrinsic line of T Tauri stars.  Our
spectrum shows the star is of spectral class A with emission in the core
of the H$\alpha$ absorption line (Figure 3i).  Spectral energy
distribution has its maximum between 2 and 12 $\mu$m.  Probably this is
a heavily reddened ($A_V$\,$\approx$\,10 mag) Herbig Ae star with a thin
disk or envelope.

\vskip1mm

{\bf SL\,101 = 2MASS J03290756+5701336}

The object is also identified as MSX G143.1521+00.4784.  It is located
in the cloud T\,942, 1.1\degr\ from GL\,490.  The image has a
north-directed 8\arcsec\ long jet-like structure.  In Paper II it was
classified as a YSO of Class III on the grounds of a MSX point at 8.3
$\mu$m.  In our spectrum a strong emission in H$\alpha$ and fainter
emissions in O\,I and Ca\,II lines are seen; its spectral type is about
G0.  According to the strength of emissions the star should be a T Tauri
star or YSO of Class II.

\vskip1mm

{\bf SL\,158 = 2MASS J03300545+5813253}

The object is also known as IRAS 03261+5803 and MSX G142.5800+01.5382.
Located in the same clump of the dust cloud T\,942 as GL\,490,
0.6\degr\ from it.  In Paper III it is classified as a YSO of Class II.
In our spectrum a strong emission in the core of absorption H$\alpha$
line is present, spectral type Ae.

\enlargethispage{3mm}

\newpage

{\bf The four Gahm stars}

As was described in Paper I, these stars have been discovered as
containing H$\alpha$ emission by Gahm (1990) in low-dispersion
objective-prism spectra, obtained with the Stockholm Observatory Schmidt
telescope.  In Paper I these four stars were selected from 12
H$\alpha$-line emission objects discovered in this area:  they lie above
or near the intrinsic line of T Tauri stars in the $J$--$H$ vs.
$H$--$K_s$ diagram.  The H$\alpha$ emission for three of them (Gahm 21,
22 and 23) was recently confirmed by the IPHAS photometric survey
(Witham et al. 2008).  The stars are located at the eastern part of the
Cam OB1 association, near the Sh\,2-205 nebula.

The blue, red and infrared SkyView images of Gahm 21 show a broad `tail'
which in the red image extends southward up to 10\arcsec.  At about
14\arcsec\ to SW, a faint star or a knot of the `tail' is seen.  In the
blue image the `tail' has an extended form running 22\arcsec\ along
the W--E direction and includes the mentioned knot.  The star Gahm 25 at
$\sim$\,6\arcmin\ southward has a much fainter neighbor seen in all
colors from $B$ to $K_s$.

In our spectra all four Gahm stars exhibit strong H$\alpha$ emissions.
The emissions of O\,I, Ca\,II and P9 lines are much fainter.
Only three of them (Gahm 21, 23 and 25) seem to be T Tauri-type stars of
spectral classes G and K. The star Gahm 22 is a heavily reddened Herbig
Ae/Be star with $E_{H-K}$\,$\approx$\,0.5 mag; this corresponds to
$A_V$\,$\approx$\,7 mag.

{\bf CGCS 475 = 2MASS J03191389+5841586}

This star located at a distance of $\sim$\,1\degr\ from GL\,490 was
included into the observing program by chance.  Later it was found that
the star is a carbon star CGCS 475 (Alksnis et al. 2001), its magnitude
$V$ seems to be close to 15.0 and variable.  Probably the star may be
identified with IRAS 03154+5831, but their positions differ by
62.6\arcsec.  The spectrum shown in Figure 3p is typical for N-type
carbon stars, with strong CN bands.  Another similar carbon star, IRAS
03156+5828, discovered from the IRAS low-resolution spectra
(Chen \& Chen 2003) is located just 6\arcmin\ left of CGCS 475. It is
much fainter in optical but brighter in $K_s$ than CGCS 475.


\sectionb{4}{CONCLUSIONS}

The far-red slit spectra of 15 known and suspected YSOs embedded in
dust/mo\-le\-cular clouds of Camelopardalis and the nearby region of
Cassiopeia are obtained.  Among them is the well-investigated infrared
object GL\,490, a pre-Herbig Ae/Be star in a dense gas and dust shell.
All these objects exhibit strong H$\alpha$ emission lines and fainter
emission lines of O\,I, Ca\,II and P9.  The spectral classes of these
objects range from A to K. Since equivalent widths of H$\alpha$
emissions for most of the objects (except one) are larger than 10 \AA,
these stars should be either in the T Tauri stage (7
objects) or in the Herbig Ae/Be stage (6 objects).  One object is a star
of intermediate temperature and one seems to be in a post T Tauri stage.
The conclusion is in agreement with the form of spectral energy
distributions of some of these stars in the medium and far infrared
(Papers II and III).  Some of the objects exhibit jets or outflows.  In
the future, the light, line intensity and spectral energy distribution
variations are to be verified.

Consequently, we confirm that the brightest objects, which in Papers II
and III have been suspected to be YSOs, indeed are in the
pre-main-sequence stage of evolution.  This fact gives a credence that
the fainter objects, selected by the same method in the mentioned
papers, should be also in the same evolutionary stage.  Recently, 13
more YSOs from Papers II and III were confirmed to be H$\alpha$ emission
stars by the IPHAS photometric survey (Witham et al. 2008;
Gonz\'alez-Solares et al. 2008).  If the majority of the suspected YSOs
will be confirmed by photometric and spectroscopic observations, the
Camelopardalis dark cloud area can be recognized as a region of
active star formation, similar to the Taurus SFR on the opposite
side of the Galactic equator.


\thanks {We are thankful to the Steward Observatory for the observing
time and to Edmundas Mei\v stas for his help preparing the paper.  The
use of the 2MASS, IRAS, MSX, IPHAS, SkyView, Gator and Simbad databases
and the IRAF program package is acknowledged.}

\References

\refb Alksnis A., Balklavs A., Dzervitis U., Eglitis I., Paupers O.,
Pundure I. 2001, {\it General Catalog of Galactic Carbon Stars by C.\,B.
Stephenson}, 3rd Edition, Baltic Astronomy, 10, 1

\refb Alonso-Costa J. E., Kwan J. 1989, ApJ, 338, 403

\refb Bunn J. C., Hoare M. G., Drew J. E. 1995, MNRAS, 272, 346

\refb Chen P.-S., Chen W.-P. 2003, AJ, 125, 2215

\refb Danks A. C., Dennefeld M. 1994, PASP, 106, 382

\refb Dobashi K., Uehara H., Kandori R., Sakurai T., Kaiden M.,
Umemoto T., Sato F. 2005, PASJ, 57, S1

\refb Dolidze M. V. 1975, Bull. Abastumani Obs., 47, 3

\refb Gonz\'alez-Solares E. A., Walton N. A., Greimel R., Drew J. E. et
al. 2008, MNRAS (in press) = arXiv 0712.0384

\refb Herbig G. H. 1962, Advances in Astron. \& Astrophys., 1, 47

\refb Herbig G. H., Bell K. R. 1988, {\it Third Catalog of Emission-Line
Stars of the Orion Population}, Lick Obs.  Bull., No. 1111

\refb Karr J. L., Martin P. G. 2003a, ApJ, 595, 880

\refb Karr J. L., Martin P. G. 2003b, ApJ, 595, 900

\refb Kerton C. R., Brunt C. M. 2003, A\&A, 399, 1083

\refb Kohoutek L., Wehmeyer R. 1997, {\it H-alpha Stars in Northern
Milky Way}, Abh. Hamburger Sternw., 11, Teil 1+2 = CDS catalog III/205

\refb McGregor P. J., Persson S. E., Cohen J. G. 1984, ApJ, 286, 609

\refb Ogura K., Sugitani K., Pickles A. 2002, AJ, 123, 2597

\refb Price S. D., Walker R. G. 1976, {\it The AFGL Four Color Infrared
Sky Survey: Catalog of Observations at 4.2, 11.0, 19.8, and 27.4
$\mu$m}, Report AFGL-TR-76-0208

\refb Simon M., Felli M., Cassar L., Fischer J., Massi M. 1983, ApJ,
266, 623

\refb Strai\v zys V., Laugalys V. 2007a, Baltic Astronomy, 16, 167
(Paper I)

\refb Strai\v zys V., Laugalys V. 2007b, Baltic Astronomy, 16, 327
(Paper II)

\refb Strai\v zys V., Laugalys V. 2008, Baltic Astronomy, 17, 1
(Paper III, this issue)

\refb Witham A. R., Knigge C., Drew J. E., Greimal R. et al. 2008,
MNRAS, 384, 1277

\end{document}